\documentclass[11pt]{amsart}
\usepackage{color}
\usepackage{amssymb,bigints,fullpage,bbm,amsaddr}
\usepackage{graphics}

\def\dtau{\partial^{(\tau)}}

\def\soft{{\rm soft}}
\def\stiff{{\rm stiff}}
\def\eff{{\rm eff}}
\def\xit{\xi^{(t)}}

\def\P{\mathcal P}
\def\Port{\mathcal P_{\bot}}

\def\e{\varepsilon}

\DeclareMathOperator*{\dom}{\mathrm{dom}}
\DeclareMathOperator*{\Ker}{\mathrm{Ker}}

\newtheorem{theorem}{Theorem}[section]

\newtheorem{definition}[theorem]{Definition}
\newtheorem{lemma}[theorem]{Lemma}

\newtheorem{remark}[theorem]{Remark}

\usepackage{amsopn}

\usepackage[ntheorem]{empheq}
\usetagform{default}

\begin{document}

\title{Time-dispersive behaviour as a feature of critical-contrast media}

\author{Kirill Cherednichenko}
\address{Department of Mathematical Sciences, University of Bath, Claverton Down, Bath, BA2 7AY, United Kingdom }
\email{K.Cherednichenko@bath.ac.uk}
\author{Yulia Ershova}
\address{Department of Mathematical Sciences, University of Bath, Claverton Down, Bath, BA2 7AY, United Kingdom {\sc and} Department of Mathematics, St.\,Petersburg State University of Architecture and Civil Engineering, 2-ya Krasnoarmeiskaya St. 4, 190005 St.\,Petersburg, Russia }
\email{julija.ershova@gmail.com}
\author{Alexander V. Kiselev}
\address{Department of Higher Mathematics and Mathematical Physics, St.\,Petersburg State University, Ulianovskaya 1, St.Peterhoff, 198504 St.\,Petersburg, Russia}
\email{alexander.v.kiselev@gmail.com}



\thanks{KDC and YE is grateful for the financial support of the Engineering and Physical Sciences Research Council: Grant EP/L018802/2 ``Mathematical foundations of metamaterials: homogenisation, dissipation and operator theory''. AVK has been partially supported by the  RFBR grant 16-01-00443-a.}

\maketitle

\begin{abstract}
Motivated by the urgent need to attribute a rigorous mathematical meaning to the term ``metamaterial'',  we propose a novel approach to the homogenisation of critical-contrast composites. This is based on the asymptotic analysis of the Dirichlet-to-Neumann map on the interface between different components (``stiff'' and ``soft'') of the medium, which leads to an asymptotic approximation of eigenmodes. This allows us to see that the presence of the soft component makes the stiff one behave as a class of time-dispersive media. By an inversion of this argument, we also offer a recipe for the construction of such media with prescribed dispersive properties from periodic composites.



\end{abstract}





\section{Introduction}

\subsection{Physics context and motivation for quantitative analysis}

Understanding the dependence of material properties of continuous media on frequency is a natural and practically relevant task, stemming from the theoretical and experimental studies of ``metamaterials'', {\it e.g.} materials that exhibit negative refraction of propagating wave packets. Indeed, it was noted as early as in the pioneering work \cite{Veselago}, that negative refraction is only possible under the assumption of frequency dispersion, {\it i.e.} when the material parameters (permittivity and permeability in electromagnetism, elastic moduli and mass density in acoustics) are not only frequency-dependent, but also become negative in certain frequency bands.

Independently of the search for metamaterials, in the course of the development of the theory of electromagnetism, it has transpired in modern physics that the Maxwell equations need to be considered with time-nonlocal ``memory'' terms, see {\it e.g.} \cite[Section 7.10]{Jackson} and also \cite{Cessenat}, \cite{Tip_1998}. The related generalised system (in the absence of charges and currents in the domain of interest) has the form
\begin{equation}
\rho\partial_tu+\int_{-\infty}^ta(t-\tau)u(\tau)d\tau+{\rm i}Au=0,\qquad A=\left(\begin{array}{cc}0 & {\rm i}\,{\rm curl}\\[0.3em] {\rm -i}\,{\rm curl}& 0\end{array}\right),
\label{gen_Maxwell}
\end{equation}
\noindent where $u$ represents the (time-dependent) electromagnetic field $(H, E)^\top$, the matrix $\rho$ depends on the electric permittivity and magnetic permeability,
and $a$ is a matrix-valued ``susceptibility" operator, set to zero in the more basic form of the system.\footnote{From the rigorous operator-theoretic point of view, $A$ in (\ref{gen_Maxwell}) is treated as a self-adjoint operator in a Hilbert space $\mathbb H$ of functions of $x\in\Omega,$ for example $\mathbb H=L^2(\Omega; {\mathbb R}^6),$ where $\Omega$ is the part of the space occupied by the medium.}

Applying the Fourier transform in time $t$ to (\ref{gen_Maxwell}), an equation in the frequency domain is obtained:
\begin{equation}
\bigl(i\omega\rho+\widehat{a}(\omega)\bigr)\widehat{u}(\cdot,\omega)+iA\widehat{u}(\cdot, \omega)=0,
\label{frequency_dep}
\end{equation}
\noindent where $\widehat{u}$ is the Fourier transform of $u,$ and $\omega$ is the frequency.
Equation (\ref{frequency_dep}) is often interpreted as a ``non-classical'' version of Maxwell's system of equations, where the permittivity and/or permeability are frequency-dependent. The existence of such media (commonly known as Lorentz materials) and the analysis of their properties go back a few decades in time and has also attracted considerable interest quite recently, {\it e.g.} in the study of plasma in tokamaks, see \cite{Weder} and references therein.


Simultaneously with the above developments in the physics literature, recent mathematical evidence, see \cite{Jikov}, \cite{Bullshitte}, suggests that  such novel material behaviour, which is incompatible (see \cite{Birman,ChKisYe_PDE,ChKisYe}) with the mathematical assumption of uniform ellipticity of the corresponding differential operators
(such as $A$ in (\ref{gen_Maxwell})), may be explained by means of the asymptotic analysis (``homogenisation'') of operator families with rapidly oscillating, and non-uniformly elliptic, coefficients.

It is therefore reasonable to ask the question of whether frequency dispersion laws such as pertaining to (\ref{frequency_dep}), which in turn may provide one with metamaterial behaviour in appropriate frequency intervals \cite{Veselago}, can be derived by some process of homogenisation of composite media with contrast (or, as we shall suggest below, any other miscroscopic degeneracies resonating with the macroscopic wavefields).

\subsection{Basis for the mathematical framework} If one were to look for an asymptotic expansion of eigenmodes of a high-contrast composite,  {\it restricted} to the soft component of the medium, one would notice (see, e.g., \cite{CherKis}) that their leading order terms can be understood as the eigenmodes of boundary-value problems with impedance (i.e., frequency dependent) boundary conditions. Such problems have been considered in the past (see, e.g., \cite{PavlovFaddeev}), motivated by the analysis of the wave equation. On the other hand, by the celebrated analysis of the so-called generalised resolvents of \cite{Naimark,Naimark1943} one knows, that a problem of this type admits a conservative dilation, which is constructed by adding the hidden degrees of freedom. In fact, precisely this latter observation has been used in \cite{Figotin_Schenker_2005,Figotin_Schenker_2007b} in devising a conservative ``extension'' of a time-dispersive system of the type \eqref{gen_Maxwell}. The substance of the argument that is proposed in the present paper is that the aforementioned conservative dilation is in fact precisely the asymptotic model of the original high-contrast composite. Furthermore, the leading order terms of its eigenmodes restricted to the {\it stiff} component are solutions to a problem of the type \eqref{frequency_dep} with frequency dispersion. They can be easily expressed in terms of the above impedance boundary value problems, thus yielding an explicit description of the link between the resonant soft inclusions and the macroscopic time-dispersive properties.

Therefore, models of continuous media with frequency-dependent effective boundary conditions can be seen as natural building blocks for media with frequency dispersion.

\begin{figure}[h!]
\begin{center}
\includegraphics[scale=1.0]{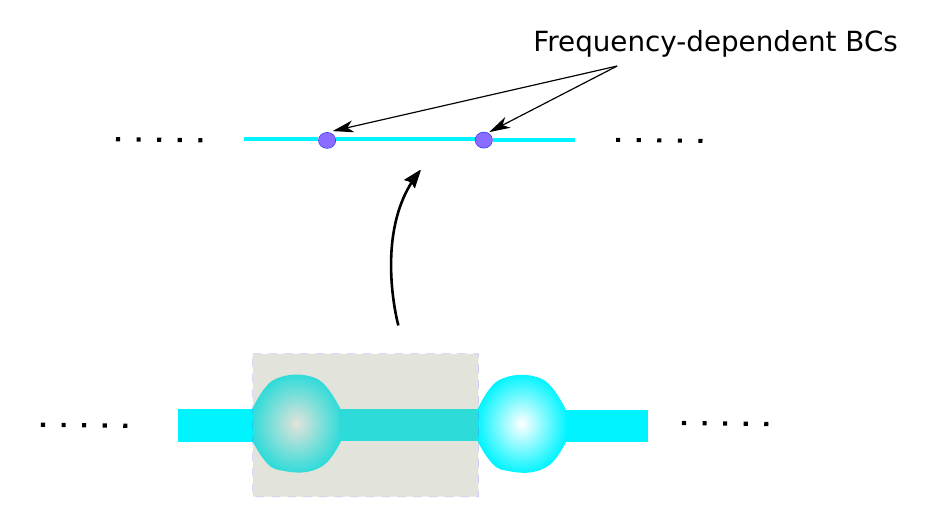}
\end{center}
\caption{{\scshape An example of a resonant thin network} {\small Edge volumes are asymptotically of the same order as vertex volumes. The stiffness of the material of the structure is of the order period-squared. }}
\label{fig:exner}
\end{figure}

It is of a considerable value to relate these ideas to the earlier works \cite{KuchmentZeng, KuchmentZeng2004, Exner}, where similar limiting impedance-type problems are obtained in the spectral analysis of ``thin" periodic structures, converging to metric graphs. Here, one obtains the aforementioned impedance setup (see Fig. \ref{fig:exner}) on the limiting graph as the asymptotics of the eigenmodes of a Neumann Laplacian, when the ``thickness'' of the structure vanishes for one particular (resonant) scaling between the ``edge'' and ``vertex''  volumes of the structure.

It is instructive to point out that the results of \cite{CherKis} establish a thrilling relationship between the analysis of thin structures and the homogenisation theory of high-contrast composites.


\begin{figure}[h!]
\begin{center}
\includegraphics[scale=1.0]{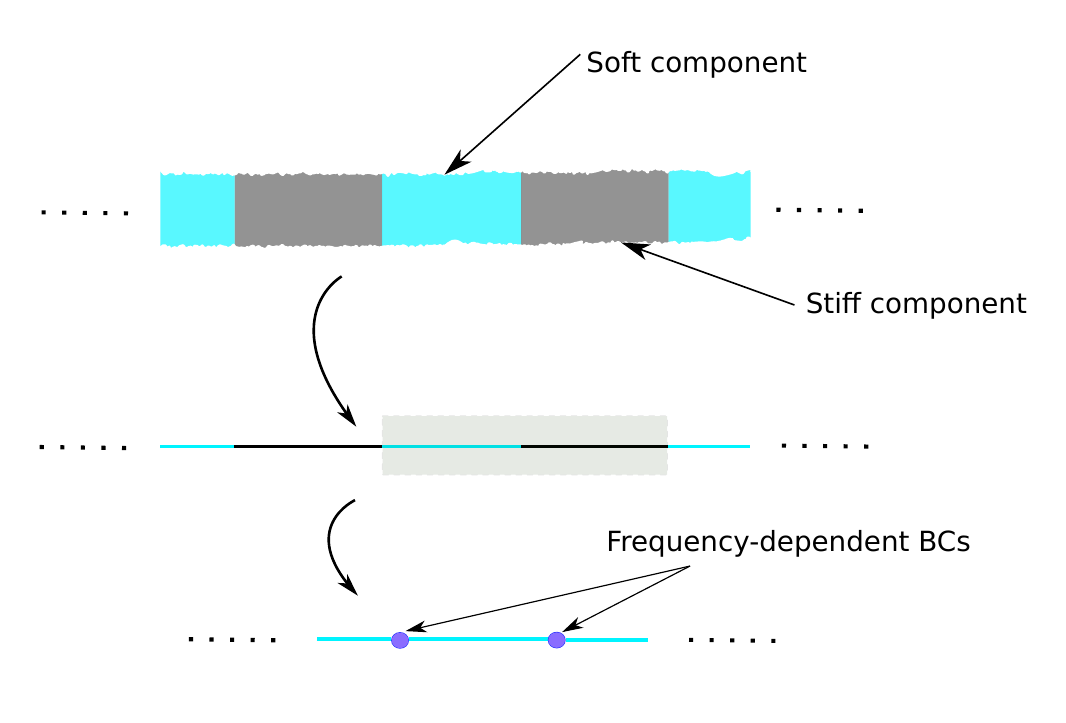}
\end{center}
\caption{{\scshape High-contrast superlattice} {\small The problem for a superlattice is reduced to a one-dimensional high-contrast problem. This is asymptotically equivalent to an impedance-type problem on the soft component.}}
\label{chain}
\end{figure}
Namely, the paper \cite{CherKis} deals with the case of the so-called superlattices \cite{Tsu} with high contrast, see Fig. \ref{chain}. 
 While simple to set up, the related system of ordinary differential equations (subject to the  appropriate conditions of continuity of fields and fluxes) is nontrivial from the point of view of quantitative analysis, see also \cite{ChCG}. It is shown that the asymptotic model for this system is precisely the one derived in \cite{KuchmentZeng, KuchmentZeng2004, Exner} in the case of a resonant thin structure converging to a chain-graph, see Fig. \ref{fig:exner}.
As we shall argue in the present article, such superlattices (and the corresponding chain-graphs) offer a simple prototype for a metamaterial, via the mathematical approach outlined above.

The result described above suggests, that thin networks might acquire the same asymptotic properties as those of the corresponding high-contrast composites. It is therefore a viable conjecture, that the metamaterial properties of a medium  can be attained via a version of geometric contrast instead of relying upon the contrast between material components. This is especially promising when the required material contrast cannot be guaranteed, as is commonly the case in elasticity and electromagnetism. The corresponding thin networks on the other hand have been made available in the study of graphenes and related areas. This subject will be further pursued in a forthcoming publication.

\begin{figure}[h!]
\begin{center}
\includegraphics[scale=0.7]{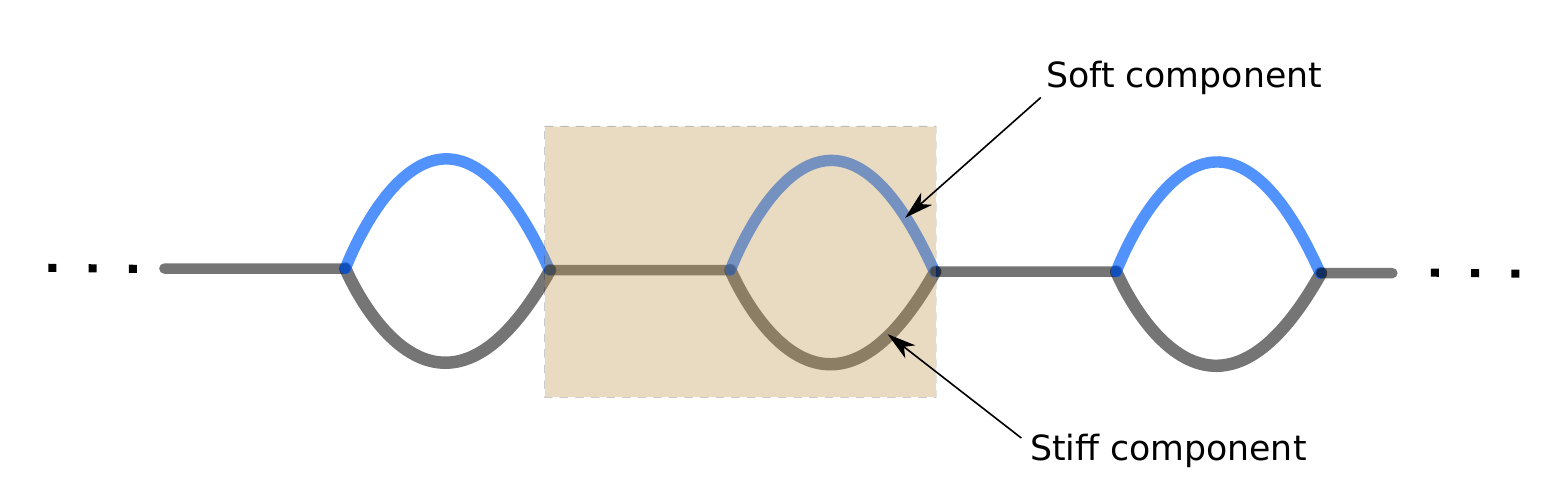}
\end{center}
\caption{{\scshape Thin network} {\small An example of a high-contrast periodic network. Stiff channels are in grey, soft channels are in blue.}}
\label{thick_chain}
\end{figure}

The above exposition vindicates the value of quantum graph models in the analysis of high-contrast composites, where we follow the well-established convention, see \cite{Kuchment2}, to use the term {\it quantum graph} for an ordinary differential operator of second order defined on a metric graph. These graph-based models are seen as natural limits of composite thin networks consisting of a large number of channels (for, say, acoustic or electromagnetic waves), where a combination of high-contrast and rapid oscillations becomes increasingly taxing at small scales and often leads to impractical numerical costs.
For channels with low cross-section-to-length ratios,
the material response of such a system, see Fig.\,\ref{thick_chain}, is closely approximated by a quantum graph as described above.
Systems of this type are a particular example of high-contrast composites and thus, as explained above, they possess resonant properties at the miscroscale, which leads to macroscopic dispersion by the above argument. At a very crude level, this  is similar to the way in which particle motion on the atomic scale leads to Lorentz-type electromagnetism, see {\it e.g.} \cite[Chapter 1]{Nussenzweig} for the analysis of a related model of damped harmonic oscillator.

Furthermore, periodic quantum graphs with vanishing period can serve as realistic explicitly solvable ODE models for multidimensional continuous media, as demonstrated\footnote{We remark, that it was Professor Pavlov who had pioneered the mathematical study of quantum graphs, see \cite{Pavlov_old}.}, e.g., in \cite{MelnikovPavlov}, where an $h-$periodic cubic lattice is shown to be close (up to and including the scattering properties) to the Laplacian in $\mathbb R^d$. More involved periodic graphs can be used to model non-trivial media, including anisotropic ones.

As a particular realistic example of a thin network with high contrast, consider the problem of modelling acoustic wave propagation in a  system of channels $\Omega^{\varepsilon, \delta}$,  $\varepsilon$-periodic in one direction, of thickness $\delta\ll\varepsilon$, and with contrasting material properties (cf. Fig. \ref{thick_chain}). To simplify the presentation, we assume the antiplane shear wave polarisation (the so called S-waves), which leads to a scalar wave equation for the only non-vanishing component $W,$ of the form
\[
W_{tt}-\nabla_x\cdot (a^\varepsilon(x)\nabla_xW)=0,\qquad u=W(x,t),\quad x, t\in{\mathbb R},
\]
where the coefficient $a^\varepsilon$  takes values one and $\varepsilon^2$ in different channels of the $\varepsilon$-periodic structure.
Looking for time-harmonic solutions $W(x,t)=U(x)\exp({\rm i}\omega t),$ $\omega>0,$ one arrives at the spectral problem
\begin{equation}
-\nabla\cdot (a^\varepsilon\nabla U)=\omega^2U.
\label{spectral}
\end{equation}
As we argue below, the behaviour of (\ref{spectral}) is close, in a quantitatively controlled way as $\varepsilon\to0,$ to that of an ``effective medium'' on ${\mathbb R}$ described by an equation of the form
\begin{equation}
-U''=\beta(\omega)U,
\label{dispersive}
\end{equation}
for an appropriate function $\beta=\beta(\omega)$, explicitly given in terms of the material parameters $a^\varepsilon$ and the topology of the original system of channels.

The goal of the present paper is to derive an explicit general formula for the function $\beta$ in (\ref{dispersive}), in terms of the topology of the graph representing the original domain of wave propagation, which is no longer restricted to the example shown in Fig.\,\ref{thick_chain}. As noted above, the presence of both rapid oscillations and high contrast make the task mathematically nontrivial. In our approach, which is new, we call upon some recently developed machinery in the operator-theoretic analysis of abstract boundary-value problems (which in our case take the form of boundary-value problems for differential operators of interest). In the subsequent work \cite{ChKisYe_PDE} we develop the corresponding analysis for the multidimensional case, which is neither included nor an extension of the analysis for graphs presented in this article. However, it is based on the same set of mathematical ideas, which makes us hope that the foundations for (\ref{dispersive}) in the case of PDEs is clear from what follows.

Unlike the approach aimed at derivation of norm-resolvent convergence, which we adopt in \cite{ChKisYe,ChKisYe_PDE}, in the present paper, having the convenience of the more physically inclined reader in mind, we systematically treat the subject from the point of view of spectral problems and in particular of the asymptotic analysis of eigenmodes. We refer the interested reader to the aforementioned papers, where further mathematical details, which we think are out of scope here, are contained.



The present paper can be viewed as following in the footsteps of \cite{CherKis} in that it relies upon the analysis of the fibre representations (obtained via the Floquet-Gelfand transform) of the original periodic operator. This is carried out using the boundary triples  theory (see, e.g., \cite{Gor,DM}), which generalises the classical methods based on the Weyl-Titchmarsh $m-$coefficient, applied to self-adjoint extensions of symmetric operators. This allows us to develop a novel approach to the homogenisation of a class of periodic high-contrast problems on ``weighted quantum graphs'', {\it i.e.} one-dimensional versions of thin composite media where the material parameters on one of the components are much lower than on the others and scaled in a ``critical'' way with respect to the period of the composite. We reiterate  that the idea that such media can be viewed as idealised models of thin periodic critical-contrast networks has been explored in the mathematics literature, see \cite{KuchmentZeng2004}, \cite{Exner}, \cite{Zhikov_singular_structures} and elsewhere. The backbone of our approach is, as explained above, the study of eigenfunctions of the problem restricted to one (``soft'') component of the composite only. After the asymptotics for these is obtained, it proves possible to reconstruct the ``complete'' eigenfunctions, where we implicitly rely upon the classical results of operator theory, in particular dealing with out-of-space self-adjoint extensions of symmetric operators and associated generalised resolvents.

\subsection {Physics interpretation and relevance to metamaterials}

Our argument leads to the understanding of the phenomenon of critical-contrast homogenisation limit as a manifestation of a frequency-converting device: if one restricts the eigenfunctions to the ``stiff'' component, they prove to be close to those of the medium where the soft component has been replaced with voids, \emph{but} correspond to non-trivially shifted eigenfrequencies. This is precisely what one would expect in the setting of time-dispersive media after the passage to the frequency domain, {\it cf.} (\ref{frequency_dep}).

From the physics perspective, this link between homogenisation and frequency conversion can be viewed as a justification of an ``asymptotic equivalence'' between eigenvalue problems for periodic composites with high contrast and problems with nonlinear dependence on the spectral parameter, which in the frequency domain characterise ``time-dispersive media'', as in (\ref{gen_Maxwell}), see also \cite{Tip_1998, Tip_2006, Figotin_Schenker_2005,
Figotin_Schenker_2007b}.

As we mentioned above, the phenomenon of frequency dispersion emerging as a result of homogenisation has been observed in the two-scale formulation applied to critical-contrast PDEs in, {\it e.g.}, \cite{Jikov, Bullshitte}. Our approach goes beyond the results of \cite{Jikov, Bullshitte} in several ways. First, being based on an explicit asymptotic analysis of operators, using the recent developments in the theory of abstract boundary-value problems (see {\it e.g.} \cite{Ryzh_spec}), it  provides an explicit procedure for recovering the dispersion relation and does not draw upon the well-known two-scale asymptotic techniques.

The approach we develop in the present paper thus offers a new perspective on frequency-dispersive (time non-local) continuous media in the sense that it provides a recipe for the construction of such media with prescribed dispersive properties from periodic composites whose individual components are non-dispersive. 
It has been known that time-dispersive media \cite{Figotin_Schenker_2005} in the frequency domain can be realised as a ``restriction'' of a conservative Hamiltonian defined on a space which adds the ``hidden'' degrees of freedom.\footnote{
This is based on the observation that the equation (\ref{frequency_dep}) can be written in the form of an eigenvalue problem ${\mathcal A}U=\omega U,$ $U\in{\mathcal H},$ for a suitable  self-adjoint ``dilation" ${\mathcal A}$ of the operator $A,$ so that ${\mathcal A}$ acts in a space ${\mathcal H}\supset{\mathbb H}.$ The vector field  $U$ has a natural physical interpretation in terms of additional electromagnetic field variables, the so-called polarisation $P$ and magnetisation $M,$ so that the full (12-dimensional) field vector is $(H, E, P, M)^\top.$ }

In summary, the existing belief in the engineering and physics literature that time-dispersive properties often arise as the result of complex microstructure of composites suggests to look for a rather concrete class of such conservative Hamiltonian dilations, namely, those pertaining to differential operators on composites with critical contrast. Our results can be viewed as laying foundations for rigorously solving this problem.

\section{Infinite-graph setup}

Consider a graph ${\mathbb G}_\infty,$ periodic in one direction, so that ${\mathbb G}_\infty+\ell={\mathbb G}_\infty,$ where $\ell$ is a fixed vector, which defines the graph axis.
Let the periodicity cell  ${\mathbb G}$ be a finite compact graph of total length $\varepsilon\in(0,1),$ and denote by
$e_j,$ $j=1,2,\dots n,$ $n\in{\mathbb N}$ its edges. For each $j=1,2,\dots, n,$ we identify $e_j$ with the interval $[0,\varepsilon l_j],$ where $\varepsilon l_j$ is the length of $e_j.$  We associate with the graph ${\mathbb G}_\infty$ the Hilbert space
$$
L_2({\mathbb G}_\infty):=\bigoplus\limits_{{\mathbb Z}}\bigoplus\limits_{j=1}^n L_2(0, \varepsilon l_j).
$$

Consider a sequence of operators $A^\varepsilon,$ $\varepsilon>0,$ in $L_2({\mathbb G}_\infty),$ generated by second-order differential expressions
\begin{equation}
-\frac{d}{dx}\left(\bigl(a^\varepsilon\bigr)^2\frac{d}{dx}\right),
\label{diff_operator}
\end{equation}
with positive ${\mathbb G}$-periodic coefficients $(a^\varepsilon)^2$ defined on ${\mathbb G}_\infty,$
with the domain ${\rm dom}(A^\varepsilon)$ that describes the coupling conditions at the vertices of
${\mathbb G}_\infty:$
\begin{equation}
{\rm dom}(A^\varepsilon)=
\left\{
u\in\bigoplus\limits_{e\in{\mathbb G}_\infty}W^{2,2}\bigl(e)\Big|\ u
\text{\ continuous,}\ \sum_{e\ni V}\sigma_e(a^\varepsilon)^2u'(V)=0\
\forall\ V\in{\mathbb G}_\infty\right\},
\label{Atau}
\end{equation}
In the formula (\ref{Atau}) the summation is carried out over the edges $e$
sharing the vertex $V,$ the coefficient $(a^\varepsilon)^2$ in the vertex condition is calculated on the edge $e,$ and $\sigma_{ e}=-1$ or $\sigma_{e}=1$ for $e$  incoming or outgoing  for $V,$ respectively. 
The matching conditions \eqref{Atau} represent the so-called standard, or Kirchhoff, conditions of combined continuity of the function and equality to zero of sums of co-normal derivatives at all vertices.

\section{Gelfand transform}
\label{Gelfand_section}
We seek to apply the one-dimensional Gelfand transform
\begin{equation}\label{1dGelfand}
v(x)=\sqrt{\frac{\varepsilon}{2\pi}}\sum\limits_{n\in \mathbb{Z}}u(x+\varepsilon n) {\rm e}^{-it(x+\varepsilon n)}.
\end{equation}
to the operator $A^\e$ defined on ${\mathbb G}_\infty$ in order to obtain the direct fibre integral for the operator $A^\varepsilon:$
\begin{equation}\label{vonNeumann}
A^{\varepsilon}=\int_{\oplus}A^\varepsilon_t dt.
\end{equation}
In order to do achieve this goal, we first note that the geometry of ${\mathbb G}_\infty$ is encoded in the matching conditions \eqref{Atau} \emph{only}. This opens up a possibility to embed the graph ${\mathbb G}_\infty$ into $\mathbb R^1$ by rearranging it edges
as consecutive segments of the real line (leading to a one-dimensional $\e$-periodic chain graph). In doing so we drop the customary practice of drawing graphs in a way reflecting matching conditions ({\it i.e.}, so that these are local relative to graph vertices). The above embedding leads to rather complex non-local matching conditions, but, on the positive side, allows us to use the Gelfand transform as required by \eqref{1dGelfand}, \eqref{vonNeumann}.

The Gelfand transform leads to periodic conditions on the boundary of the cell $\mathbb G$ and thus in our case identifies the ``left" boundary vertices of the graph $\mathbb G$ with their translations by $\ell$, which results in a modified graph $\widehat{\mathbb G}$. Apart from this, the matching conditions for the internal vertices of $\mathbb G$ admit the same form as for $A^\e$, except for the fact that the Kirchhoff matching is replaced by a Datta-Das Sarma one (the latter can be viewed as a weighted Kirchhoff), see below in \eqref{Atau1}. Unimodular weights appearing in Datta-Das Sarma conditions are precisely due to the non-locality of matching conditions mentioned above for the embedding of $\mathbb G_\infty$ into $\mathbb R^1$.

The image of the Gelfand transform is described as follows. There exists a unimodular list $\{w_V(e)\}_{e\ni V},$ {\it cf.} \cite{ChKisYe}, defined at each vertex $V$ of $\widehat{\mathbb G}$ as a finite collection of values corresponding to the edges adjacent to $V$.
For each $t\in[-\pi/\varepsilon,\pi/\varepsilon)$, the fibre operator $A^\varepsilon_t$  is generated by the differential expression
\begin{equation}
\left(\frac 1i \frac d{dx}+t\right)(a^\varepsilon)^2\left(\frac 1i \frac d{dx}+t\right)
\label{diff_expr}
\end{equation}
on the domain
\begin{multline}
{\rm dom}(A^\varepsilon_t)=
\Bigg\{
v\in\bigoplus\limits_{e\in {\mathbb G}}W^{2,2}\bigl(e)\ \Big|\ w_V(e)v|_e(V)=w_V(e')v|_{e'}(V)
\text{\ for all } e,e' \\ \text{ adjacent to } V, \ \sum_{e\ni V}\partial^{(t)}v(V)=0\ \ \
{\rm for\ each\ vertex}\ V\Bigg\},
\label{Atau1}
\end{multline}
where $\partial^{(t)}v(V)$ is the weighted ``co-derivative''  $\sigma_{e}w_V(e)(a^\varepsilon)^2(v'+{\rm i}t v)$  of the function $v$ on the edge $e,$ calculated at $V.$



\section{Boundary triples for extensions of symmetric operators}
\label{triples_section}

In the analysis of the asymptotic behaviour of the fibres of the original operator representing the quantum graph, we employ the framework of boundary triples for a symmetric operator with equal deficiency indices for the description of a class of its extensions. Part of the toolbox of the theory of boundary triples is the generalisation of the classical Weyl-Titchmarsh $m$-function to the case of a matrix (finite deficiency indices) and operators (infinite deficiency indices).

The boundary triples theory is a very convenient toolbox for dealing with extensions of linear operators, originating in the works of M.\,G. Kre\u\i n. In essence, it is an operator-theoretic interpretation of the second Green's identity. As such, it allows one to pass over from the consideration of functions in Hilbert spaces to a formulation in which one deals with objects in the boundary spaces (such as traces of functions and traces of their normal derivatives), which in the context of quantum graphs are finite-dimensional. Furthermore, it allows one to use explicit concise formulae for the resolvents of operators under scrutiny and for other related objects. Thus it facilitates the analysis by expressing the familiar, commonly used in this area, objects in a concise way.


\begin{definition}[\cite{Gor,Ko1,DM}]
Suppose that $A_{\rm max}$ is the adjoint to a densely defined symmetric operator on a separable Hilbert space $H$ and let $\Gamma_0,$ $\Gamma_1$ be linear mappings of ${\rm dom}(A_{\max})\subset H$
to a separable Hilbert space $\mathcal{H}.$

A. The triple
$(\mathcal{H}, \Gamma_0,\Gamma_1)$ is called \emph{a boundary
triple} for the operator $A_{\max}$  
if the following two conditions hold:
\begin{enumerate}
\item For all $u,v\in {\rm dom}(A_{\max})$ one has the second Green's identity
\begin{equation}
\langle A_{\max} u,v \rangle_H -\langle u, A_{\max} v \rangle_H = \langle \Gamma_1 u, \Gamma_0
v\rangle_{\mathcal{H}}-\langle\Gamma_0 u, \Gamma_1 v\rangle_{\mathcal{H}}.
\label{Green_identity}
\end{equation}
\item The mapping
${\rm dom}(A_{\max})\ni u\longmapsto (\Gamma_0 u,
\Gamma_1 u)\in{\mathcal H}\oplus{\mathcal H}$
is onto.
\end{enumerate}

B. A restriction ${A}_B$ of the operator $A_{\rm max}$ such
that $A_{\rm max}^*=:A_{\min}\subset  A_B\subset A_{\max}$  is called
almost solvable if there exists a boundary triple
$(\mathcal{H}, \Gamma_0,\Gamma_1)$ for $A_{\max}$ and a bounded
linear operator $B$ defined on $\mathcal{H}$ such that
\[
{\rm dom}({A_B})=\bigl\{u\in{\rm dom}(A_{\rm max}):\ \Gamma_1u=B\Gamma_0u\bigr\}.
\]

C. The operator-valued Herglotz\footnote{For a definition and properties of Herglotz functions, see {\it e.g.} \cite{Nussenzweig}.} function $M=M(z),$ defined by
\begin{equation}
\label{Eq_Func_Weyl}
M(z)\Gamma_0 u_{z}=\Gamma_1 u_{z}, \ \
u_{z}\in \ker (A_{\max}-z),\  \ z\in
\mathbb{C}_+\cup{\mathbb C}_-,
\end{equation}
is called the Weyl-Titchmarsh $M$-function of the operator
$A_{\max}$ with respect to the corresponding boundary triple.
\end{definition}


Suppose $A_B$ be a self-adjoint almost solvable restriction of $A_{\rm max}$ with compact resolvent. Then $M(z)$ is analytic on the real line away from the eigenvalues of $A_\infty,$ where $A_\infty$ is the restriction of $A_{\rm max}$ to domain $\dom(A_\infty)=\dom(A_{\rm max})\cap\ker(\Gamma_0).$ It is a key observation for what follows that $u\in{\rm dom}(A_B)$ is an eigenvector of $A_B$ with eigenvalue $z_0\in{\mathbb C}\setminus{\rm spec}(A_\infty)$ if and only if
\begin{equation}
\bigl(M(z_0)-B\bigr)\Gamma_0u=0.
\label{eigeneq}
\end{equation}


In the next section we utilise a particular operator $A_{\rm max}$ and a boundary triple $({\mathcal H}, \Gamma_0, \Gamma_1),$ which we use to analyse the resolvents of the operators on quantum graphs
introduced earlier.



\section{Graph with high contrast: prototype for time-dispersive media}
\label{our_graph}
In what follows we develop a general approach to the analysis of weighted quantum graphs with critical contrast. We demonstrate it on one particular example, which, as we show in Appendix A, exhibits all the properties of the  generic case. We have therefore chosen to present the analysis in the terms that are immediately applicable to the general case and, whenever advisable, we provide statements that carry over without modifications. Speaking of a ``general'' case, we imply an operator of the class introduced in Section 2, where some of the edges $e_{\text{soft}}$ of the cell graph $\mathbb G$ carry the weight $a^\e=\e$, with the remaining edges carrying weights of order 1 uniformly in $\e$.

The rationale of the present section is in fact extendable to an even  more general setup (including the one of periodic high-contrast PDEs), which we treat in the paper \cite{ChKisYe_PDE}. However, in the present paper we consider a rather simplified model, in view of keeping technicalities to a bare minimum and thus hopefully making the matter transparent to the reader.

\begin{figure}[h!]
\begin{center}
\includegraphics[scale=1.5]{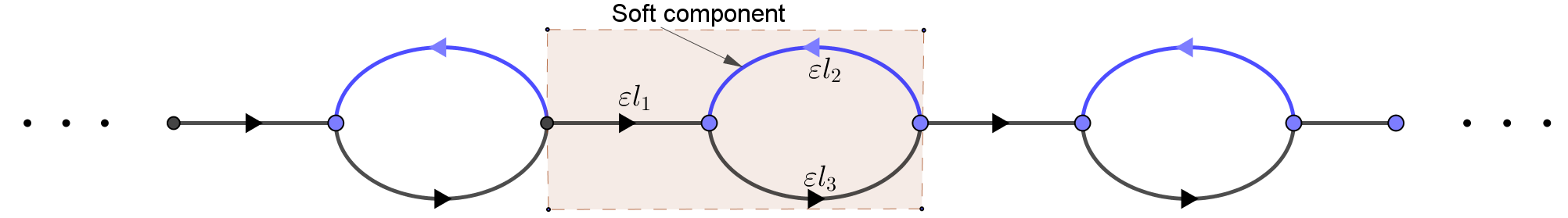}
\end{center}
\caption{{\scshape Periodicity cell $\mathbb G.$} {\small The intervals of lengths $\varepsilon l_1$ and $\varepsilon l_3$ are ``stiff", {\it i.e.} they carry the weights $a_1^2$ and $a_3^2$, respectively, whereas the interval of length $\varepsilon l_2$ is ``soft", with weight $\varepsilon^2.$}}
\label{infinite_graph_figure}
\end{figure}

\begin{figure}[h!]
\begin{center}
\includegraphics[scale=0.7]{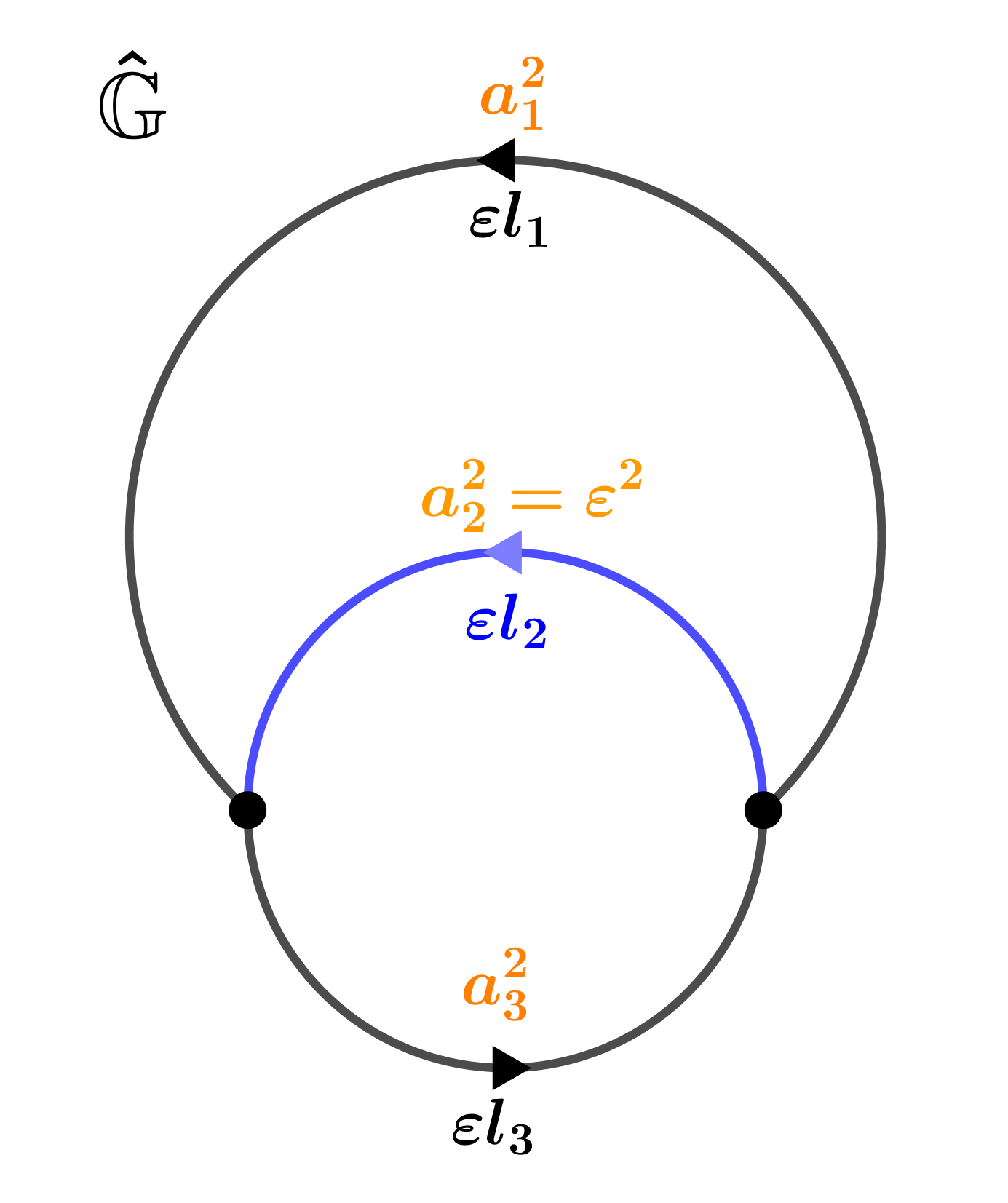}
\end{center}
\caption{{\scshape The graph $\widehat {\mathbb G}.$} {\small The left and right boundary vertices have been identified.}}
\label{compact_fig}
\end{figure}

Consider the graph ${\mathbb G}_\infty$ with the periodicity cell ${\mathbb G}$ shown in Figure  \ref{infinite_graph_figure}.
The Gelfand transform, see Section \ref{Gelfand_section}, applied to this graph, yields the graph $\widehat{\mathbb G}$ of Figure \ref{compact_fig}. In the present section we show that there exists a boundary triple such that $A^\e_t$ is an almost solvable extension of the corresponding $A_{\min}$, and the
$M$-function (which is in our case a matrix-valued function; for convenience, it is written as a function of $k:=\sqrt{z}$, with the branch chosen so that $\Im k>0$) of $A_{\max}$ is given by
\begin{equation}
\label{Msplit}
M(k,\varepsilon, t)=k\widetilde{M}^{\rm stiff}(\varkappa,\tau)+\varepsilon\widetilde{M}^{\rm soft}(k, \tau),\quad \varkappa:=\varepsilon k,\quad\tau:=\varepsilon t,
\end{equation}
where
\[
\widetilde{M}^{\rm stiff}(\varkappa,\tau):=\left(\begin{array}{cc}
-a_1\cot\dfrac{\varkappa l_1}{a_1} -a_3\cot\dfrac{\varkappa l_3}{a_3}\ &\
a_1\dfrac{{\rm e}^{-i(l_1+l_3)\tau}}{\sin\dfrac{\varkappa l_1}{a_1}}+a_3 \dfrac{{\rm e}^{il_2\tau}}{\sin\dfrac{\varkappa l_3}{a_3}}\\[3.3em]
a_1\dfrac{{\rm e}^{i(l_1+l_3)\tau}}{\sin\dfrac{\varkappa l_1}{a_1}}+a_3 \dfrac{{\rm e}^{-il_2\tau}}{\sin\dfrac{\varkappa l_3}{a_3}}\ &\
-a_1\cot\dfrac{\varkappa l_1}{a_1}-a_3\cot\dfrac{\varkappa l_3}{a_3}
\end{array}\right),
\]
\begin{equation}
\widetilde{M}^{\rm soft}(k,\tau):=k\left(\begin{array}{cc}-\cot k l_2\ &\ \dfrac{{\rm e}^{il_2\tau}}{\sin k l_2}\\[1.6em]
\dfrac{{\rm e}^{-il_2\tau}}{\sin k l_2}\ &\ -\cot k l_2
\end{array}\right),
\label{M_soft_tilde}
\end{equation}

Note that  for all $\tau\in[-\pi, \pi),$ the function $\widetilde{M}^{\rm soft}(\cdot,\tau)$ is meromorphic and regular at zero.


Essentially, the claim made is a straightforward consequence of the double integration by parts, followed by a simple rearrangement of terms.
In the rest of this section we sketch the construction applicable in the general case, which in particular yields the above claim for the model graph considered. Under the definitions of Section \ref{triples_section}, the maximal operator
$A_{\rm max}=A_{\rm min}^*$ is defined by the same differential expression (\ref{diff_expr}) on the domain
\begin{multline}\label{domAmax}
{\rm dom}(A_{\rm max})=
\biggl\{
v\in\bigoplus\limits_{e\in \widehat{\mathbb G}}W^{2,2}\bigl(e)\ \Big|\ w_V(e)v|_e(V)=w_V(e')v|_{e'}(V)
\\
\text{\ for all } e,e' \text{ adjacent to } V,\ \
\forall\,V \in\widehat{\mathbb G}
\biggr\}.
\end{multline}
In what follows we use the triple $({\mathbb C}^m, \Gamma_0, \Gamma_1),$ where $m$ is the number of vertices in the graph $\widehat{\mathbb G}$, and
\begin{equation}
\Gamma_0v=\bigl\{v(V)\bigr\}_V,
\qquad \Gamma_1v=\Bigl\{\sum_{e\ni V}\partial^{(t)}v(V)\Bigr\}_V,\qquad v\in{\rm dom}(A_{\rm max}),
\label{boundary_operators}
\end{equation}
where $u(V)$ is defined as the common value of  $w_V(e)v|_e(V)$ for all edges $e$ adjacent to $V$.

By definition of the $M$-matrix one has
$
\Gamma_1v=M\Gamma_0v,$
$
v\in\ker (A_{\rm max} - z).
$
Functions $v\in\ker (A_{\rm max} - z)$
have the form
$$
v(x)=\exp(-{\rm i}xt)\biggl\{A_e\exp\biggl(-\frac{{\rm i}kx}{a^\varepsilon}\biggr)+B_e\exp\biggl(\frac{{\rm i}kx}{a^\varepsilon}\biggr)\biggr\},\quad x\in e,\quad A_e, B_e\in{\mathbb C},
$$
where $k:=\sqrt{z}$, and the co-derivative is given by
$$
a_\e^2(v'(x)+it v(x))
= {\rm i}ka^\varepsilon\exp(-{\rm i}xt)
\biggl\{-A_e\exp\biggl(-\frac{{\rm i}kx}{a^\varepsilon}\biggr)+B_e\exp\biggl(\frac{{\rm i}kx}{a^\varepsilon}\biggr)\biggr\}, \qquad x\in e,
$$
For the vertex $V$
and for every ``Dirichlet data" vector $\Gamma_0v$ one of whose entries is unity and the other entries vanish,
the ``Neumann data'' vector $\Gamma_1v$ gives the column of the $M$-matrix corresponding to $V.$
The corresponding $\Gamma_1v$ has diagonal and off-diagonal entries of the form, respectively,
$$
-\sum_{e\in V} ka^\varepsilon\cot\left(\dfrac{k \varepsilon l_e}{a^\varepsilon}\right),\qquad
\qquad \sum_{e\in V} ka^\varepsilon \widetilde w_V(e)\biggl(\sin \dfrac{k \varepsilon l_e}{a^\varepsilon}\biggr)^{-1},
$$
where $\{\widetilde w_V(e)\}_{e\ni V}$ is a unimodular list uniquely determined by the list $\{w_V(e)\}_{e\ni V}$.
The resulting $M$-matrix is constructed from these columns over all vertices $V.$

In particular, for the example of Fig. \ref{infinite_graph_figure} -- \ref{compact_fig}, we have the following:
the boundary space $\mathcal H$ pertaining to the graph $\widehat{\mathbb G}$ is chosen as $\mathcal H=\mathbb C^2$. The unimodular list functions $w_{V_1}$ and $w_{V_2}$ are chosen as follows:
\begin{equation*}\label{eq:1-weights}
\begin{gathered}
\{w_{V_1}(e^{(j)})\}_{j=1}^3=\{1,1,e^{i\tau(l^{(2)}+l^{(3)})}\},\quad \{w_{V_2}(e^{(j)})\}_{j=1}^3=\{e^{i\tau l^{(3)}},1,1\}
\end{gathered}
\end{equation*}
and
\begin{equation*}
\begin{gathered}
\{\widetilde w_{V_1}(e^{(j)})\}_{j=1}^3=\{e^{-i\tau(l^{(1)}+l^{(3)})},e^{i\tau l^{(2)}},e^{i\tau l^{(2)}}\},\\ \{\widetilde w_{V_2}(e^{(j)})\}_{j=1}^3=\{e^{i\tau(l^{(1)}+l^{(3)})},e^{-i\tau l^{(2)}},e^{-i\tau l^{(2)}}\},
\end{gathered}
\end{equation*}
yielding the formula \eqref{M_soft_tilde}.



\section{Asymptotic diagonalisation of the $M$-matrix and the eigenvector asymptotics}\label{sect:asymp_diag}

The present section is the centrepiece of our approach. The major difficulty to overcome is of course the fact that the operator $A^\varepsilon_t$ entangles in a non-trivial way the stiff and soft components of the medium. On the level of the analysis of the operator itself this problem admits no obvious solution, unless one is prepared to introduce a two-scale asymptotic ansatz. On the other hand, the $M$-matrix calculated above will be shown to be additive with respect to the decomposition of the medium (hence the notation $M^{\rm soft}$ and $M^{\rm stiff}$). Thus, via the representation \eqref{eigeneq}, it proves possible to use the asymptotic expansion of $M^{\rm stiff}$, which is readily available, to recover the asymptotics of eigenmodes, restricted to the soft component. This way, the homogenisation task at hand can be viewed as a version of the perturbation analysis in the boundary space pertaining to the problem.

In the example considered (and in the general case in view of Appendix A)
it follows from (\ref{eigeneq}), (\ref{Msplit})
that $u_\varepsilon$ is an eigenfunction of the operator $A^\varepsilon_t,$ see (\ref{diff_expr})--(\ref{Atau1}), if and only if
\begin{equation}
M^{\rm soft}\Gamma_0u_\varepsilon=-M^{\rm stiff}\Gamma_0u_\varepsilon,\qquad M^{\rm soft}:=\varepsilon\widetilde{M}^{\rm soft},\quad
M^{\rm stiff}:=k\widetilde{M}^{\rm stiff}.
\label{bc1}
\end{equation}
In writing (\ref{bc1}), we assume, without loss of generality, that the eigenvalue $z_\varepsilon=k^2$ corresponding to the eigenfunction $u_\varepsilon$ does not belong to the spectrum of the Dirichlet decoupling $A_\infty^t,$ defined according to the general theory of Section \ref{triples_section} for the operators we introduce in Section \ref{Gelfand_section}. Indeed, in any compact subset of $\mathbb C,$ for small enough $\varepsilon,$ this spectrum coincides with the $\varepsilon$-independent set of poles of the matrix
$\widetilde{M}^\soft,$ see (\ref{M_soft_tilde}). For the same reason, we can safely assume that the eigenvalues $z_\varepsilon$ do not belong to the spectrum of the Dirichlet operator on the soft inclusion. These assumptions ensure that that the condition $z_0\in{\mathbb C}\setminus{\rm spec}(A_\infty)$ for the validity of (\ref{eigeneq}) is satisfied in both cases: for the $M$-matrix of the operator $A^\varepsilon_t,$ where $B=0,$ and for the $M$-matrix of the operator on the soft component represented by (\ref{bc1}), where the role of $B$ is played by the matrix $-M^\stiff.$

We proceed by observing that the matrices $M^{\text{soft}}$ and $M^{\text{stiff}}$ in (\ref{bc1}) can be treated as $M$-matrices of certain triples on their own. In particular, it will be instrumental in what follows to attribute this meaning to $M^{\text{soft}}$. To this end, consider the decomposition of the graph $\widehat{\mathbb G}$ into its ``soft'' $\mathbb G^{\text{soft}}$ and ``stiff'' $\mathbb G^{\text{stiff}}$ components (each of these is treated as a graph, so that $\widehat{\mathbb G}=\mathbb G^{\text{soft}}\cup \mathbb G^{\text{stiff}}$) and the operator $A_{\max}^{\text{soft}}$ defined by \eqref{diff_expr}, \eqref{domAmax}, with $\widehat{\mathbb G}$ replaced by $\mathbb G^{\text{soft}}$. The boundary space for $A_{\max}^{\soft}$ can be defined as $\mathcal H$, the same as the boundary space for the operator $A_{\max}$ (again by Appendix A in the general case). The boundary operators $\Gamma_j^{\text{soft}}$, $j=0,1,$ are defined as in \eqref{boundary_operators} for the graph $\mathbb G^{\text{soft}}$. Then, by inspection, the $M$-matrix for the operator $A_{\max}^{\text{soft}}$ is nothing but $M^{\text{soft}}$ (see \cite{CherKisSilva} for further details).

For each $v\in{\rm dom}(A_{\rm max}),$ define $\widetilde{v}$ to be the restriction of $v$ to the soft component $\mathbb G^{\text{soft}}$. It is obvious that $\widetilde{v}\in\dom(A_{\max}^{\soft}).$

We notice that (\ref{bc1}) implies, in particular, that
\begin{equation}
M^{\rm soft}\Gamma_0^{\rm soft}\widetilde{u}_\e=B^\varepsilon\Gamma_0^{\rm soft}\widetilde{u}_\e,\qquad\qquad B^\varepsilon:=-M^{\rm stiff}.
\label{suggested_bound_cond}
\end{equation}
Furthermore, since $M^{\rm soft}$ is the $M$-matrix for the pair $(\Gamma^{\rm soft}_0, \Gamma^{\rm soft}_1),$ one has
\[
M^{\rm soft}\Gamma_0^{\rm soft}\widetilde{u}_\e=\Gamma_1^{\rm soft}\widetilde{u}_\e,
\]
so the condition (\ref{suggested_bound_cond}) takes a form similar to (\ref{Eq_Func_Weyl}):
\begin{equation}
\Gamma_1^{\rm soft}\widetilde{u}_\varepsilon=B^\varepsilon\Gamma_0^{\rm soft}\widetilde{u}_\varepsilon.
\label{eq_eigenvector3}
\end{equation}

This condition involves the Dirichlet data of the solution to the spectral equation for $A_{\max}^{\text{soft}}$ which is an ODE on the graph $\mathbb G^{\text{soft}}$ with a constant coefficient. The Dirichlet data $\Gamma_0^{\rm soft}\widetilde{u}_\varepsilon$ determine the vector $\widetilde{u}_\e$ uniquely. The named vector is interpreted as a solution to the spectral equation on the soft component of the graph $\widehat{\mathbb G}$ subject to
$z$-dependent boundary conditions, encoded in \eqref{eq_eigenvector3}. On the other hand, this vector can also be used to reconstruct the vector $u_\e$: indeed, from $\Gamma_0 u_\e=\Gamma_0^{\text{soft}}\widetilde{u}_\e$ it follows, that $u_\e$, which is by assumption an eigenvector to $A^\e_t$ at the point $z$, is nothing but a continuation of $\widetilde{u}_\e$ to the rest of the graph $\widehat{\mathbb G}$ based on its Dirichlet data at the boundary of the soft component. It follows, cf. \eqref{eq_eigenvector3}, that the asymptotic analysis can be reduced to the soft component, with the information about the presence of the stiff component fed into the related asymptotic procedure by means of the stiff-soft interface.

Before we proceed further, let us take another look at the equation $M\Gamma_0u_\e=0,$ {\it cf.} (\ref{bc1}), which is equivalent to $u_\e$ being an eigenvector of $A^\e_t$ at the value of spectral parameter $z$. Using the fact that $M=M^{\text{soft}}+M^{\text{stiff}}$ as well as the explicit expressions for the matrices $M^{\text{soft}},$ $M^{\text{stiff}},$ {\it cf.} \eqref{Msplit}, it is easily seen that the leading-order term of $\Gamma_0 u_\e$, and thus of $u_\e$, does not depend on the soft component of the medium, since the elements of $M^{\text{soft}}$ are $\e$-small. On the other hand, the situation is drastically different from the viewpoint of the associated dispersion relation, which must be guaranteed for the \emph{solvability} of $M \Gamma_0 u_\e=0$. The dispersion relation follows from the condition $\det M=0$, and it is \emph{here, and here only}, that the soft component of the medium makes its presence felt in the problem. Due to the fact that $M^{\text{stiff}}$ is rank one at $\tau=0$, it transpires that the leading-order term of the equation $\det M=0$ \emph{in the case of critical contrast only} blends together in a non-trivial way the stiff and soft components of the medium. Bearing this in mind, the phenomenon of critical-contrast homogenisation can be seen as a manifestation of a frequency-converting device: if one restricts the eigenfunctions to the stiff component, they are $\e$-close to those of the medium where the soft component has been replaced with voids, \emph{but} correspond to non-trivially shifted eigenfrequencies. This is precisely what one would expect in the setting of time-dispersive media after the passage to the frequency domain, {\it cf.} (\ref{gen_Maxwell}), (\ref{frequency_dep}). We will come back to this discussion in Section 8.

Let us return to the analysis of \eqref{eq_eigenvector3}, which, as explained above, contains all the information on the asymptotic behaviour of $A^\e_t$. We notice that the named equation corresponds to a homogeneous ODE; the non-trivial dependence on $\e$ is concealed in the right-hand side, which describes $\e$- \emph{and} frequency-dependent boundary conditions. The problem of asymptotic analysis of eigenfunctions of $A^\e_t$ is thus effectively reduced to the analysis of the asymptotic behaviour of these boundary conditions. This analysis however is greatly simplified by the fact that $B^\e$ is equal to $-M^{\text{stiff}}$, where $M^{\text{stiff}}$ is shown to be the $M$-matrix of $A_{\max}^{\text{stiff}}$ (see Appendix A) by a similar argument to that applied above to $M^{\soft}$. Hence, the asymptotics sought for $M^{\stiff}$ is simply the asymptotics of the Dirichlet-to-Neumann map of a uniformly elliptic problem at zero frequency, which allows to use well-known elliptic techniques.

Firstly, we notice that the results of Section 5 combined with the asymptotic formulae
\begin{equation*}
a_e \cot \frac{\varkappa l_e}{a_e} = \frac{a_e^2}{\varkappa l_e}-\frac 13\varkappa l_e+ O(\varkappa^3),\quad\quad\quad a_e\biggl(\sin\dfrac{\varkappa l_e}{a_e}\biggr)^{-1} = \frac{a_e^2}{\varkappa l_e}+\frac{1}{6}\varkappa l_e+ O(\varkappa^3),
 \end{equation*}
yield the following statement.


\begin{lemma}
\label{M_expansion}
Suppose that $K\subset{\mathbb C}$ is compact.
One has
\begin{equation*}
\widetilde{M}^{\rm stiff}(\varkappa, \tau)
=\varkappa^{-1}M_0(\tau)+\varkappa M_1(\tau)+O(\varkappa^3),\quad \tau\in[-\pi, \pi),\ \varkappa=\varepsilon k,\ \varepsilon\in(0,1),\ k\in K,
\end{equation*}
where $M_0$ and $M_1$ are analytic matrix functions of $\tau$.
\end{lemma}

It follows from Lemma \ref{M_expansion} that, for all  $\tau\in[-\pi,\pi),$
\begin{equation}
B^\varepsilon(z)=\varepsilon^{-1}B_0
+\varepsilon zB_1+O(\varepsilon^3z^2),\qquad\varepsilon\in(0,1),\ \sqrt{z}\in K,
\label{asympt}
\end{equation}
where $B_0,$ $B_1$ are Hermitian matrices that depend on $\tau$ only.
The following two lemmata carry over to the general case with minor modifications, since they only pertain to the stiff component of the medium and therefore rely upon the general uniformly elliptic properties of the latter.
\begin{lemma}
\label{mu_lemma}
There exist $\gamma\geq0$ (where $\gamma=0$ if and only if the graph $\mathbb{G}^{\rm stiff}$  is a tree\footnote{Recall that a tree is a connected forest \cite{Cvetkovich}.}) and an eigenvalue branch $\mu^{(\tau)}$ for the matrix $B_0,$ such that
$\dim\Ker\bigl(B_0-\mu^{(\tau)}\bigr)=1,$ $\tau\in[-\pi,\pi),$ and
\begin{equation}
\mu^{(\tau)}=\gamma\tau^2 + O(\tau^4).
\label{mu_asymptotics}
\end{equation}
\end{lemma}

We denote by $\psi^{(\tau)}$ the normalised eigenvector for the eigenvalue $\mu^{(\tau)},$ so that
 $\psi^{(0)}=(1/\sqrt{2})(1,1)^\top,$ {\it i.e.} the trace of the first eigenvector of the Neumann problem on the stiff component at zero quiasimomentum, which is clearly constant. 
  Let $\mathcal P:=\langle \cdot, \psi^{(\tau)}\rangle \psi^{(\tau)}$ and $\mathcal P_{\bot}$ be the orthogonal projections in the boundary space onto $\psi^{(\tau)}$ and its orthogonal complement, respectively.

\begin{lemma}
\label{bound_below_lemma}
There exists $C_\perp>0$ such that
\begin{equation}
\Port B_0 \Port\ge C_\perp\Port,
\label{bound_below}
\end{equation}
in the sense that the operator $\Port(B_0-C_\perp)\Port$ is non-negative.
\end{lemma}

We use Lemma \ref{bound_below_lemma} to solve \eqref{eq_eigenvector3} asymptotically. The overall idea is to diagonalise the leading order term $\varepsilon^{-1} B_0$ of the asymptotic expansion of $B^\varepsilon$ in \eqref{eq_eigenvector3}. From Lemma \ref{mu_lemma} we infer that $B_0$ has precisely one eigenvalue quadratic in $\tau$ (which thus gets close to zero), while Lemma \ref{bound_below_lemma} provides us with a bound below on the remaining eigenvalue. The fact that the eigenvalue $\mu^{(\tau)}$ degenerates requires that the next term in the asymptotics of $B^\varepsilon$ be taken into account in the related eigenspace. This additional term is easily seen to be $z-$dependent (in fact, linear in $z$).

We start with an auxiliary rescaling of the soft component. Namely, we introduce the unitary operator
$\Phi_\varepsilon$ mapping $v\mapsto \widehat{v}$ according to the formula $\widehat{v}(\cdot)=\sqrt{\e}{v}(\e\cdot)$. Under this mapping, the length of the soft component loses its dependence on $\varepsilon$. The operator $\widehat{A}_{\rm max}^{\text{soft}}$ is defined as the unitary image of $A_{\rm max}^{\text{soft}}$ under the mapping $\Phi_\varepsilon$, and $\widehat{\Gamma}_0^\soft,$ $\widehat{\Gamma}_1^\soft$ are the boundary operators for the rescaled soft component:
\[
\widehat{\Gamma}_0^{\rm soft}\widehat{v}:=\bigl\{\widehat{v}(V)\bigr\}_V,
\qquad \widehat{\Gamma}_1^{\rm soft}\widehat{v}:=\biggl\{\sum_{e\ni V}\widehat{\partial}^{(\tau)}\widehat{v}(V)\biggr\}_V,\qquad \widehat{v}\in {\rm dom}\bigl(\widehat{A}_{\rm max}^{\text{soft}}\bigr),
\]
where we set $\widehat{v}(V)$ as the common value of  $w_V(e)\widehat{v}|_e(V)$ for all $e$ adjacent to $V,$ and
$\widehat{\partial}^{(\tau)} \widehat{v}(V)$ is the expression  $\sigma_{e}w_V(e)(\widehat{v}'+{\rm i}\tau \widehat{v})$  on the edge $e,$ calculated at $V.$ Note that $\widehat{\Gamma}_1^{\rm soft}$ does not depend on $\varepsilon$.

Under the rescaling $\Phi_\varepsilon$ the equation \eqref{eq_eigenvector3} becomes
\begin{equation}
\widehat{\Gamma}_1^{\rm soft}\widehat{u}_\varepsilon=\varepsilon^{-1} B^\varepsilon\widehat{\Gamma}_0^{\rm soft}\widehat{u}_\varepsilon,
\label{eq_eigenvector_resc}
\end{equation}
where in accordance with the above convention $\widehat{u}_\varepsilon=\Phi_\varepsilon \widetilde{u}_\varepsilon$.

We start our diagonalisation procedure by considering the non-degenerate ei\-gen\-spa\-ce of $B^\varepsilon$.
Applying $\Port$ to both sides of \eqref{eq_eigenvector_resc}, replacing $B^\varepsilon$ by its asymptotics (\ref{asympt}) and using (\ref{bound_below}) yields
\begin{equation}
\Port \widehat{\Gamma}_1^\soft \widehat{u}_\varepsilon =\varepsilon^{-2}\Port B_0 \Port \widehat{\Gamma}_0^\soft \widehat{u}_\varepsilon + O(1)\ge \varepsilon^{-2}C_\perp\Port \widehat{\Gamma}_0^\soft \widehat{u}_\varepsilon + O(1),
\label{asym_rel}
\end{equation}
where we assume that $u_\varepsilon$ is $L^2$-normalised.
Multiplying by $\e^2$ both sides of (\ref{asym_rel}) and applying the Sobolev embedding theorem to the left-hand side of (\ref{asym_rel}),
we infer
\begin{equation}\label{eq_part-solution}
\Port \widehat{\Gamma}_0^\soft \widehat{u}_\varepsilon = O(\e^2).
\end{equation}
Plugging this partial solution back into \eqref{eq_eigenvector_resc}, to which $\P$ is applied on both sides, we obtain
\begin{align*}
\P \widehat{\Gamma}_1^\soft \widehat{u}_\varepsilon
&= \varepsilon^{-2}\P B_0 \P \widehat{\Gamma}_0^\soft \widehat{u}_\varepsilon +  z\P B_1\P \widehat{\Gamma}_0^\soft \widehat{u}_\varepsilon + O(\e^2)\\[0.4em]
&=\varepsilon^{-2}\mu^{(\tau)}\P\widehat{\Gamma}_0^\soft \widehat{u}_\varepsilon +  z\P B_1\P \widehat{\Gamma}_0^\soft \widehat{u}_\varepsilon + O(\e^2).
\end{align*}

We have proved that up to an error term admitting a uniform estimate $O(\varepsilon^2)$ one has the following asymptotically equivalent problem for the eigenvector $\widehat{v}_\varepsilon$:
\begin{equation}\label{eq:eigenvector_asymp}
  \Port \widehat{\Gamma}_0^\soft \widehat{u}_\varepsilon =0,\quad
  \P \widehat{\Gamma}_1^\soft \widehat{u}_\varepsilon = \varepsilon^{-2}\mu^{(\tau)}\P\widehat{\Gamma}_0^\soft \widehat{u}_\varepsilon +  z\P B_1\P \widehat{\Gamma}_0^\soft \widehat{u}_\varepsilon.
\end{equation}

We use Lemma \ref{mu_lemma} and expand $\P B_1 \P$ in powers of $\tau=\varepsilon t$ as follows\footnote{In the example considered in the present paper, as opposed to the general case, one can prove that $\P B_1 \P= \P B_1^{(0)}\P+O(\tau^2)$, see the calculation in \cite[Appendix B]{ChKisYe} for details. This yields the error bound $O(\e^2)$ in the statement of Theorem \ref{eff_thm} below.}: $\P B_1 \P= \P B_1^{(0)}\P+O(\tau)$.
The second equation in \eqref{eq:eigenvector_asymp} admits the form
\begin{equation}
\label{eq_eigenvector4}
\P \widehat{\Gamma}_1^\soft \widehat{u}_\e= \gamma t^2 \P \widehat{\Gamma}_0^\soft \widehat{u}_\e+z \P B_1^{(0)}\P \widehat{\Gamma}_0^\soft \widehat{u}_\e +
(O(\tau)+O(\tau^4/\varepsilon^2))\P \widehat{\Gamma}_0^\soft \widehat{u}_\e.
\end{equation}
Expressing $\P \widehat{\Gamma}_0^\soft \widehat{u}_\e$ from the latter equation, it is easily seen based on embedding theorems that \eqref{eq_eigenvector4} is asymptotically equivalent, up to an error uniformly estimated as $O(\varepsilon)$, to the following equation:
\begin{equation}
\label{eq_eigenvector5}
\P \widehat{\Gamma}_1^\soft \widehat{u}_\e= \gamma t^2 \P \widehat{\Gamma}_0^\soft \widehat{u}_\e+z \P B_1^{(0)}\P \widehat{\Gamma}_0^\soft \widehat{u}_\e.
\end{equation}

   We formulate the above result as the following theorem.

\begin{theorem}
\label{eff_thm}
Let $\widehat{u}$ solve the  following equation on the re-scaled soft component:
\begin{equation*}\label{eq:eq}
\begin{aligned}
\widehat{A}_{\rm max}^{\rm soft}\widehat{u}(x)&=z\widehat{u}(x),\\[0.3em]
\Port \widehat{\Gamma}_0^\soft \widehat{u} &= 0,\\[0.3em]
\P \widehat{\Gamma}_1^\soft \widehat{u}&= \gamma t^2 \P \widehat{\Gamma}_0^\soft \widehat{u}+z \P B_1^{(0)}\P \widehat{\Gamma}_0^\soft \widehat{u}.
\end{aligned}
\end{equation*}

Then the eigenvalues $z_\varepsilon$ and their corresponding eigenfunctions $u_\varepsilon$ of the operators $A^\varepsilon_t$
are $O(\varepsilon)$-close uniformly in $t\in[-\pi/\varepsilon, \pi/\varepsilon)$, in the sense of $\mathbb C$ and in the sense of the $L^2$ norm,  respectively, to the values $z$ as above and functions ${u}_{\rm eff}$ defined as follows. On the soft component ${\mathbb G}^\soft$ we set $u_{\rm eff}(\cdot):=(1/\sqrt{\e})\widehat{u}({\e}^{-1}\cdot)$. On the stiff component ${\mathbb G}^\stiff$ the function $u_{\rm eff}$ is obtained as the extension by $(1/\sqrt{\e})v,$
where $v$ is the solution of the operator equation
\[
A_{\rm max}^\stiff v=0,
\]
determined by the Dirichlet data of $\widehat{u}(\varepsilon^{-1}\cdot),$ where $A_{\rm max}^\stiff$ is defined by (\ref{maternoe_slovo}), Appendix A.




\begin{remark}
It is straightforward to see that the eigenvalue $\mu^{(\tau)}$ in Lemma \ref{mu_lemma} is the least, by absolute value, Steklov eigenvalue of $A_{\max}^\stiff$, {\it i.e.} the least $\kappa$ such that the problem
$$
\begin{aligned}
A_{\max}^{\stiff}\breve v&=0, \quad \breve v\in W^{2,2}(\mathbb G^\stiff),\\[0.3em]
\Gamma_1^\stiff\breve v&=\kappa \Gamma_0^\stiff\breve v.
\end{aligned}
$$
admits a non-trivial solution $\breve v.$  Note that for this solution $\breve v$ one has $\Gamma_0^\stiff\breve v=\psi^{(\tau)}.$ It follows that for the function $v$ of Theorem \ref{eff_thm} one has $v=c\breve v,$ where $c$ is a constant determined by $\widehat{u}.$
\end{remark}



\end{theorem}




\section{Eigenvalue and eigenvector asymptotics in the example of Section \ref{our_graph}}

Here we provide the result of an explicit calculation applying the general procedure described in the previous section to the specific example of Section \ref{our_graph} (see \cite{ChKisYe} for details).
We start by expanding the matrix $B^\varepsilon$
as a series in powers of $\e$:
$$
\widehat{B}:=\varepsilon^{-1}B^\varepsilon=
\widehat{B}_0+z\widehat{B}_1+O(\varepsilon^2z^2),\
\widehat{B}_0:= { \frac{1}{\e^2 }\begin{pmatrix} D&\overline{\xi}\\[0.7em]
                                                     \xi& D
                                                     \end{pmatrix}},\ \widehat{B}_1:=
                                                     {\begin{pmatrix}  E & \overline{\eta}\\[0.7em]
                                                    \eta& E
                                                     \end{pmatrix}},
$$
where
\begin{align}
\xi:&=-\frac{a_1^2}{l_1}\exp\bigl({\rm i}\tau(l_1+l_3)\bigr)-\frac{a_3^2}{l_3}\exp(-{\rm i}\tau l_2),\quad\quad\quad D:=\frac{a_1^2}{l_1}+\frac{a_3^2}{l_3},
\label{xi_def}\\[0.75em]
\eta:&=\dfrac{1}{6}\Bigl(l_1\exp\bigl({\rm i}\tau(l_1+l_3)\bigr)+l_3\exp(-{\rm i}\tau l_2)\Bigr),\quad\quad\quad E:=\dfrac{1}{3}(l_1+l_3).\nonumber
\end{align}
 The matrix $\e^2 \widehat{B}_0$ is Hermitian and has two distinct eigenvalues, $\mu=D-|\xi|$ and $\mu_\bot=D+|\xi|$. The eigenvalue branch $\mu$ is singled out by the condition $\mu\vert_{\tau=0}=0$. In order to diagonalise the matrix $\widehat{B}_0$, consider the normalised eigenvectors $\psi^{(\tau)}=(1/\sqrt{2})(1,-\xi/|\xi|)^\top$ and $\psi^{(\tau)}_\bot=(1/\sqrt{2})(1,\xi/|\xi|)^\top$ corresponding to the eigenvalues $\mu$ and $\mu_\bot$, respectively,  and the matrix $X:=\bigl(\psi^{(\tau)}, \psi^{(\tau)}_\bot\bigr).$
The projections ${\mathcal P},$ ${\mathcal P}_\bot$ introduced in the previous section are as follows:
\[
{\mathcal P}=\frac{1}{2}\left(\begin{array}{cc}1&\dfrac{\overline{\xi}}{\vert\xi\vert}\\[1.3em] \dfrac{\xi}{\vert\xi\vert} &1\end{array}\right),\quad\quad {\mathcal P}_\bot=\frac{1}{2}\left(\begin{array}{cc}1 &-\dfrac{\overline{\xi}}{\vert\xi\vert}\\[1.3em] -\dfrac{\xi}{\vert\xi\vert} & 1\end{array}\right).
\]

It follows by a straightforward calculation that the effective spectral problem
is given by
\begin{equation}
-\biggl(\frac d{dx}+{\rm i}\tau\biggr)^2u=zu,
\label{res_eq}
\end{equation}
\begin{multline}
u(0)=-\frac {\overline{\xi}}{|\xi|}u(l_2),\\
(u'+{\rm i}\tau u)(0)+\frac {\overline{\xi}}{|\xi|}(u'+{\rm i}\tau u)(l_2)=\Biggl(\biggl(\dfrac{l_1}{a_1^2}+\dfrac{l_3}{a_3^2}\biggr)^{-1}\biggl(\frac{\tau}{\e}\biggr)^2-(l_1+l_3)z\Biggr)u(0),
\label{res_bc}
\end{multline}

By invoking Theorem \ref{eff_thm}, the problem (\ref{res_eq})--(\ref{res_bc}) on the scaled soft component provides the asymptotics, as $\varepsilon\to0,$ of the eigenvalue problems for the family $A^\varepsilon_t,$ $t=\tau/\e\in[-\pi/\varepsilon,\pi/\varepsilon).$  Its spectrum, {\it i.e.} the set of values $z$ for which (\ref{res_eq})--(\ref{res_bc}) has a non-trivial solution, as well as the corresponding eigenfunctions approximate, up to terms of order $O(\varepsilon^2),$ the corresponding spectral information for the family $A^\varepsilon_t,$ and consequently, $A^\varepsilon.$ Notice that the stiff component of the original graph (where the eigenfunctions converge to a constant, in a suitable scaled sense), appears in this limit problem through the boundary datum $u(0).$ In the next section we show that an appropriate extension of the function space for (\ref{res_eq})--(\ref{res_bc}) by the (one-dimensional) complementary space of constants leads to an eigenvalue problem for a self-adjoint operator, describing a conservative system. Solving this latter eigenvalue problem for the element in the complementary space yields a frequency-dispersive formulation we announced in the introduction.



\section{Frequency dispersion in a ``complementary" medium}

\subsection{Self-adjoint out-of-space extension}

\label{Ahom}

Following the strategy outlined at the end of the last section, we treat $u(0)$ in (\ref{res_bc}) as an additional field variable, and reformulate (\ref{res_eq})--(\ref{res_bc}) as an eigenvalue problem in a space of pairs $(u, u(0)),$ see (\ref{spectral_eq}).

More precisely, for all values $\tau\in[-\pi, \pi),$ consider an operator $A^{\rm hom}_\tau$ in the space $L^2(0, l_2)\oplus \mathbb{C}$
defined as follows. The domain $\text{\rm dom}\bigl(A^{\rm hom}_\tau\bigr)$ consist of all pairs $(u,\beta)$ such that $u\in W^{2,2}(0, l_2)$ and the quasiperiodicity condition
\begin{equation}
u(0)=\overline{w_\tau}u(l_2)=:\frac{\beta}{\sqrt{l_1+l_3}},\qquad w_\tau\in{\mathbb C},
\label{quasi_cond}
\end{equation}
is satisfied.  On $\text{\rm dom}\bigl(A^{\rm hom}_\tau\bigr)$
the action of the operator is set by
\begin{equation}
A^{\rm hom}_\tau\left(\begin{matrix}u\\[0.3em] \beta\end{matrix}\right)=
\left(\begin{array}{c}\biggl(\dfrac{1}{\rm i}\dfrac{d}{dx}+\tau\biggr)^2u\\[1.1em]
\dfrac{1}{\sqrt{l_1+l_3}}\Gamma_\tau\left(\begin{matrix}u\\[0.3em] \beta\end{matrix}\right)
\end{array}\right),
\label{lim_form}
\end{equation}
where $\Gamma_\tau: W^{2,2}(0, l_2)\oplus{\mathbb C}\to{\mathbb C}$ is bounded. We set
\begin{equation}
\Gamma_\tau\left(\begin{matrix}u\\[0.3em] \beta\end{matrix}\right)=-(u'+{\rm i}\tau u)(0)+\overline{w_\tau}
(u'+{\rm i}\tau u)(l_2)+
\frac{(\sigma t)^2}{\sqrt{l_1+l_3}}\beta,  \quad\sigma^2:=\biggl(\dfrac{l_1}{a_1^2}+\dfrac{l_3}{a_3^2}\biggr)^{-1},
\label{Gamma_part}
\end{equation}
where $w_\tau=-{\xi}/{|\xi|}$ (see \eqref{xi_def} for the definition of $\xi$), in which case $A^{\rm hom}_\tau$ is a self-adjoint operator on the domain described by (\ref{quasi_cond}). Moreover, (\ref{res_eq})--(\ref{res_bc}) is the problem on the first component of spectral problem for the operator $A^{\rm hom}_\tau:$
\begin{equation}
A^{\rm hom}_\tau\left(\begin{matrix} u\\[0.3em] \beta\end{matrix}\right)=z\left(\begin{matrix} u\\[0.3em] \beta\end{matrix}\right).
\label{spectral_eq}
\end{equation}

We now re-write this spectral problem in terms of the complementary component $\beta\in{\mathbb C}.$ In order to do this, we represent the function
$u$ in (\ref{spectral_eq}) as a sum of two: one of them is a solution to the related inhomogeneous Dirichlet problem, while the other takes care of the boundary condition. More precisely, consider the solution $v$ to the problem
\begin{equation*}
-\biggl(\frac{d}{dx}+{\rm i}\tau\biggr)^2v=0,\qquad\qquad
v(0)=1,\ \ \ \ \ v(l_2)=w_\tau,
\end{equation*}
{\it i.e.}
\begin{equation}
v(x)=\Bigl\{1+l_2^{-1}\Bigl(w_\tau\exp({\rm i}\tau l_2)-1\Bigr)x\Bigr\}\exp(-{\rm i}\tau x),\quad\quad x\in(0, l_2).
\label{function_v}
\end{equation}
The function
\[
\widetilde{u}:=u-\frac{\beta}{\sqrt{l_1+l_3}}v
\]
satisfies
\begin{equation*}
-\biggl(\frac{d}{dx}+{\rm i}\tau\biggr)^2\widetilde{u}-z\widetilde{u}=\frac{z\beta}{\sqrt{l_1+l_3}}v,\quad\qquad
 \widetilde{u}(0)=\widetilde{u}(l_2)=0.
\end{equation*}
In other words, one has
\begin{equation*}
\widetilde{u}=\frac{z\beta}{\sqrt{l_1+l_3}}(A_{\rm D}-zI)^{-1}v,
\end{equation*}
where $A_{\rm D}$ is the Dirichlet operator in $L^2(0, l_2)$ associated with the differential expression
\[
-\biggl(\frac{d}{dx}+{\rm i}\tau\biggr)^2.
\]
We can now write the ``boundary''  part of the spectral equation (\ref{spectral_eq}) as
\begin{equation}
K(\tau, z)\beta=z\beta,\quad K(\tau, z):=\dfrac{1}{l_1+l_3}\left\{z\Gamma_\tau\left(\begin{matrix}
(A_{\rm D}-zI)^{-1}v\\[0.3em] 0\end{matrix}\right)+
\Gamma_\tau\left(\begin{matrix}v\\[0.3em] \sqrt{l_1+l_3}\end{matrix}\right)\right\}.
\label{K_expr}
\end{equation}
In accordance with the rationale for introducing the component $\beta,$ the effective dispersion relation for the operator $A_{\tau/\varepsilon}^\varepsilon,$
$\tau\in[-\pi,\pi),$ is given by
\[
K(\tau, z)=z.
\]
The explicit expression for this relation that we have obtained, see (\ref{K_expr}), is new, and it quantifies explicitly the r\^ole of the soft component of the composite in the macroscopic frequency-dispersive properties. In particular, the expression (\ref{K_expr}) shows that the soft inclusions enter the macroscopic equations via a Dirichlet-to-Neumann map on the boundary of the inclusions.

\subsection{Explicit formula for the time-dispersion kernel}

Here we compute explicitly the kernel $K(\tau, z)$ entering the effective dispersion relation for $A_\tau^\varepsilon.$ In view of possible generalisations, and recalling the pioneering formula in \cite[Section 8]{Jikov} for effective dispersion in double-porosity media, we represent the action of the resolvent $(A_{\rm D}-zI)^{-1}$ as a series in terms of the normalised eigenfunctions
\begin{equation}
\phi_j(x)=\sqrt{\frac{2}{l_2}}\exp(-{\rm i}\tau x)\sin\frac{\pi jx}{l_2},\qquad x\in(0,l_2),\qquad\qquad j=1,2,3,\dots,
\label{function_phi}
\end{equation}
of the operator $A_{\rm D}.$ This yields
\begin{equation}
K(\tau, z):=\dfrac{1}{l_1+l_3}\left\{z\sum_{j=1}^\infty\dfrac{\langle v, \phi_j\rangle}{\mu_j-z}\Gamma_\tau\left(\begin{matrix}
\varphi_j\\[0.3em] 0\end{matrix}\right)+
\Gamma_\tau\left(\begin{matrix}v\\[0.3em] \sqrt{l_1+l_3}\end{matrix}\right)\right\}.
\label{K_general1}
\end{equation}
where $\mu_j=(\pi j/l_2)^2,$ $j=1,2,3,\dots,$ are the eigenvalues corresponding to (\ref{function_phi}). For the choice (\ref{Gamma_part}) of $\Gamma_\tau$ we obtain (see (\ref{function_v}), (\ref{function_phi}))
\[
\Gamma_\tau\left(\begin{matrix}v\\[0.3em] \sqrt{l_1+l_3}\end{matrix}\right)
=\frac{2}{l_2}\bigl(1-\Re\theta(\tau)\bigr)+\biggl(\frac{\sigma\tau}{\varepsilon}\biggr)^2,\qquad \theta(\tau):=\frac{\dfrac{a_1^2}{l_1}{\rm e}^{-{\rm i}\tau}+\dfrac{a_3^2}{l_3}}{\biggl|\dfrac{a_1^2}{l_1}{\rm e}^{-{\rm i}\tau}+\dfrac{a_3^2}{l_3}\biggr|},
\]
\[
\Gamma_\tau\left(\begin{matrix}
\varphi_j\\[0.3em] 0\end{matrix}\right)=-\sqrt{\frac{2}{l_2}}\frac{\pi j}{l_2}\bigl((-1)^{j+1}\overline{\theta(\tau)}+1\bigr),\
\langle v, \phi_j\rangle=\frac{\sqrt{2l_2}}{\pi j}\bigl((-1)^{j+1}\theta(\tau)+1\bigr),\ j=1,2,\dots
\]
Substituting the above expressions into (\ref{K_general1}) and making use of  the formulae, see {\it e.g.} \cite[p.\,48]{Gradshteyn_Ryzhik},
\begin{equation*}
\sum_{j=1}^\infty\frac{1}{(\pi j)^2-x^2}=\frac{1}{2}\biggl(\frac{1}{x^2}-\frac{\cos x}{x\sin x}\biggr),\quad
\sum_{j=1}^\infty\frac{(-1)^j}{(\pi j)^2-x^2}=\frac{1}{2}\biggl(\frac{1}{x^2}-\frac{1}{x\sin x}\biggr),\quad x\notin\pi{\mathbb Z},
\end{equation*}
we obtain

\begin{equation}
K(\tau, z)=\frac{1}{l_1+l_3}\biggl\{
\frac{2\sqrt{z}\cos(l_2\sqrt{z})}{\sin(l_2\sqrt{z})}
-\frac{2\sqrt{z}}{\sin(l_2\sqrt{z})}\Re\theta(\tau)+\biggl(\frac{\sigma\tau}{\varepsilon}\biggr)^2\biggr\}.
\label{K_example}
\end{equation}


\subsection{Asymptotically equivalent model on the real line}

In this section we are going to treat (\ref{K_expr}), (\ref{K_example}) as a nonlinear eigenvalue problem in the space of second components of pairs
$(u, \beta)= L^2(0, l_2)\oplus{\mathbb C}.$ As is evident from above, this problem is closely related to (\ref{res_eq})--(\ref{res_bc}), via the construction presented in Section \ref{Ahom}.
We show next that the aforementioned macroscopic field is governed by a certain frequency-dispersive formulation. In order to obtain the latter,
we will use a suitable inverse Gelfand transform.

Our strategy can be seen as motivated by the following elementary observation, closely linked with the Birman-Suslina study of homogenisation in the moderate contrast case, albeit understood in terms of spectral equations. Starting with the spectral problem
\begin{equation}\label{eq:b-s-problem}
-\frac {d^2u}{dx^2}=zu \text{\ \ on\ \ } L_2(\mathbb R),
\end{equation}
one applies the Gelfand transform\footnote{Recall, {\it cf.} Section \ref{Gelfand_section}, that the Gelfand transform is a map
$L^2({\mathbb R})\to L^2\bigl((0, \varepsilon)\times(-\pi/\varepsilon,\pi/\varepsilon)\bigr)$ given by
\begin{equation*}
{\mathcal G}u(y, t)=\sqrt{\frac{\varepsilon}{2\pi}}\sum_{n\in{\mathbb Z}}u(x+\varepsilon n)\exp\bigl(-{\rm i}t(x+\varepsilon n)\bigr),\qquad t\in\bigl[-\pi/\varepsilon, \pi/\varepsilon\bigr),\qquad x\in(0,\varepsilon).
\end{equation*}} (well-defined on generalised eigenvectors due to the rigging procedure, see, {\it e.g.,} \cite{Berezansky,BS}) to obtain for $\widetilde u:=\mathcal G u$
$$
-\biggl(\frac d {dx}+i t\biggr)^2 \widetilde u(x,t) = z \widetilde u(x,t), \quad x\in(0,\e),\quad t\in[-\pi/\e,\pi/\e).
$$
We compute the inner products of both sides in $L_2(0,\e)$ with the normalised constant function $(1/\sqrt{\e})\mathbbm 1$, which yields the dispersion relation of the original problem via the equation
$$
t^2 \widehat u (t)=z \widehat u(t),
$$
where $\widehat u$ is the Fourier transform of the function $u\in L_2(\mathbb R)$. The latter equation is then solved in the distributional sense,
\begin{equation}\label{eq:beta}
\beta (t)=\sum_{m} c_m \delta(t-t_m),
\end{equation}
where $\beta (t):=\widehat u(t)$ and the sum in \eqref{eq:beta} is taken over $m=1,2$,  $t_1, t_2$ being the zeroes of the equation $t^2=z$ and $c_m$ are arbitrary constants. Ultimately, one applies the inverse Gelfand transform
\[
({\mathcal G}^*f)(x)=\sqrt{\frac{\varepsilon}{2\pi}}\int\limits_{-\pi/\varepsilon}^{\pi/\varepsilon}f(t)\exp({\rm i}tx)dt,\quad f\in L^2\biggl(-\frac{\pi}{\varepsilon}, \frac{\pi}{\varepsilon}\biggr),\qquad x\in{\mathbb R},
\]
to the function $\mathfrak B (x,t):=(1/\sqrt{\e})\beta(t)\mathbbm 1(x),$ {\it i.e.}
$$
v(x):=\sqrt{\frac{\e}{2\pi}}\int_{-\pi/\e}^{\pi/\e} \mathfrak B(x,t) \exp(i t x) dt, \qquad x\in{\mathbb R}.
$$
It is easily seen that this function is precisely the solution to \eqref{eq:b-s-problem}.

We emulate the above argument for the case of interest to us, starting from the eigenvalue problem
$K(\tau, z)\beta=z\beta,$ which we now treat as an equation in the distributional sense with $K$ given by (\ref{K_example}). It admits  the form
\begin{equation}
(\sigma t)^2\beta=\biggl\{(l_1+l_3)z-\frac{2\sqrt{z}\cos(l_2\sqrt{z})}{\sin(l_2\sqrt{z})}+\frac{2\sqrt{z}}{\sin(l_2\sqrt{z})}\Re \theta(\varepsilon t)\biggr\}\beta,\qquad t=\frac{\tau}{\varepsilon},
\label{spectral_final}
\end{equation}
The solution is defined by \eqref{eq:beta}, where $\{t_m\}$ is the set of zeroes of the equation $K(\e t,z)=z$.

Second, we argue that the function $\mathfrak B(x,t)$ as  defined above
is the $\varepsilon$-periodic Gelfand transform of the solution to a spectral equation on ${\mathbb R}$ for a differential operator with constant coefficients, where the conventional spectral parameter $z$ is replaced by a nonlinear in $z$ expression,  as on the right-hand side of (\ref{spectral_final}).

Indeed, expand the function $\Re\theta(\tau)$ into Fourier series
\[
\Re\theta(\tau)=\frac{1}{\sqrt{2\pi}}\sum_{n=-\infty}^\infty c_n\exp({\rm i}n\tau),\qquad
c_n:=\frac{1}{\sqrt{2\pi}}\int_{-\pi}^{\pi}\Re\theta(\tau)\exp(-{\rm i}n\tau)d\tau,\qquad n\in{\mathbb Z}.
\]
and apply to $\mathfrak B(x,t)$  the inverse
Gelfand transform ${\mathcal G}^*:$
\[
({\mathcal G}^*f)(x)=\sqrt{\frac{\varepsilon  }{2\pi}}\int\limits_{-\pi/\varepsilon}^{\pi/\varepsilon}f(t)\exp({\rm i}tx)dt,\quad f\in L^2\biggl(-\frac{\pi}{\varepsilon  }, \frac{\pi}{\varepsilon  }\biggr),\qquad x\in{\mathbb R}.
\]
We denote $U:={\mathcal G}^*\mathfrak B$ and notice that
\[
\sqrt{\frac{\varepsilon  }{2\pi}}\int\limits_{-\pi/\varepsilon}^{\pi/\varepsilon}t^2\mathfrak B(x,t)\exp({\rm i}tx)dt=-\frac{d^2}{dx^2}\Biggl(\sqrt{\frac{\varepsilon  }{2\pi}}\int\limits_{-\pi/\varepsilon}^{\pi/\varepsilon}\mathfrak B(x,t)\exp({\rm i}tx)dt\Biggr)=-U''(x)
\]
and
\begin{align*}
&\sqrt{\frac{\varepsilon  }{2\pi}}\int\limits_{-\pi/\varepsilon}^{\pi/\varepsilon}\Re\theta(\varepsilon t)\mathfrak B(x,t)\exp({\rm i}tx)dt=\sum_{n=-\infty}^\infty c_n{\frac{\sqrt{\varepsilon}  }{2\pi}}\int\limits_{-\pi/\varepsilon}^{\pi/\varepsilon}\mathfrak B(x,t)\exp\bigl({\rm i}t(x+\varepsilon   n)\bigr)dt\\
\\
&=\frac{1}{\sqrt{2\pi}}\sum_{n=-\infty}^\infty c_n U(x+\varepsilon   n)
\sim
\frac{1}{\sqrt{2\pi}}\sum_{n=-\infty}^\infty c_n U(x)
=\Re\theta(0) U(x)=U(x),\qquad \varepsilon\to0.
\end{align*}

The above asymptotics as $\varepsilon\to0$ is understood in the sense of $W^{-2,2}(\mathbb R).$ It can be demonstrated, see \cite{ChKisYe}, that the order of convergence is $O(\varepsilon^{2})$ (and $O(\varepsilon)$ in the general case), however we do not dwell on the complete proof here. The idea of the proof, which is standard, can be, for example, the following. Instead of the function $\beta,$ define $\beta^0$ by the expression  (\ref{eq:beta}), where the sequence $\{t_m\}$ is replaced by the sequence $\{t_m^0\}$ of zeros of the equation $K^0(\tau, z)=z.$ Here $K^0$ is defined by (\ref{K_example}) with $\Re\theta(\tau)$ replaced by $\Re\theta(0)=1.$ It is then shown that $\beta$ is $O(\varepsilon^{2})$-close, in the sense of distributions, to $\beta^0,$ from where one obtains the claim by taking the inverse Gelfand transform of the function ${\mathfrak B}^0(x, t)=(1/\sqrt{\varepsilon})\beta^0(t){\mathbbm 1}(x).$

\begin{figure}[h!]
\begin{center}
\includegraphics[scale=1]{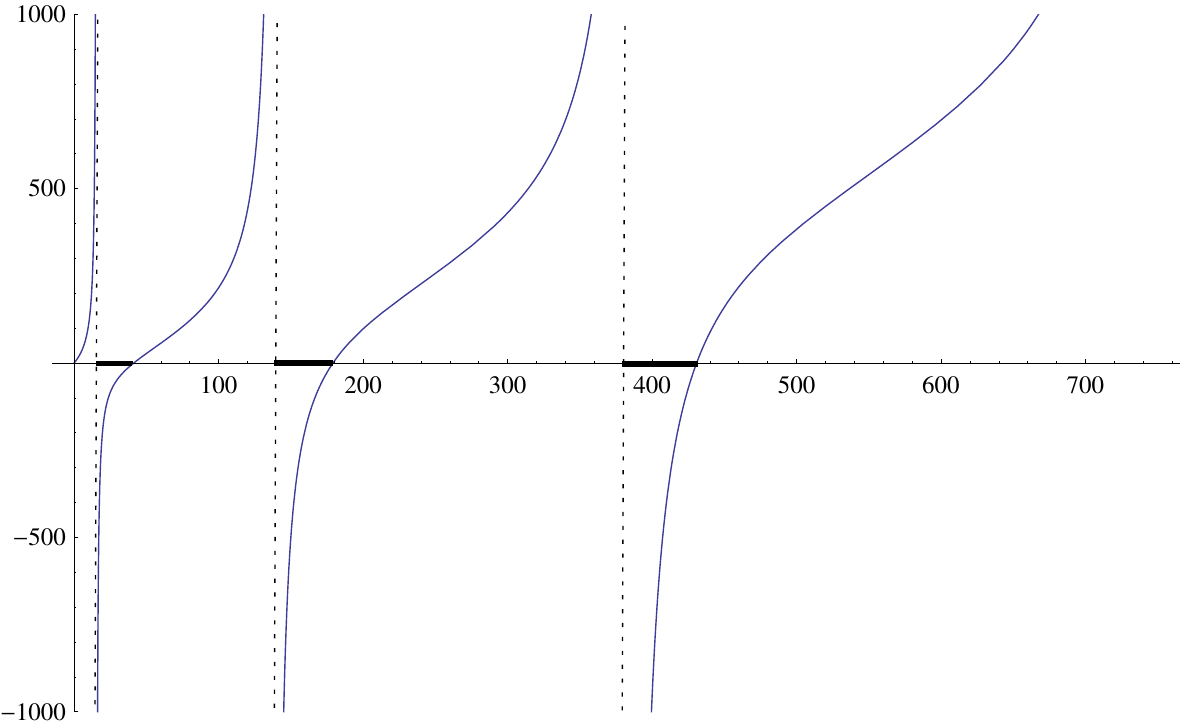}
\end{center}
\caption{{\scshape Dispersion function.} {\small The plot of the dispersion function on the right-hand side of (\ref{limit_spectral}), for $L=0.2.$ The spectral gaps are highlighted in bold.}}
\label{fig:tangens}
\end{figure}

It follows that the limit equation on the  function $U$ takes the form
\begin{equation}
-\sigma^2 \,U''(x)
=\biggl\{(l_1+l_3)z+2\sqrt{z}\tan\biggl(\frac{l_2\sqrt{z}}{2}\biggr)\biggr\}U(x),\qquad x\in{\mathbb R}.
\label{limit_spectral}
\end{equation}
In particular, the limit spectrum is given by the set of $z\in{\mathbb R}$ for which the expression in brackets on the right-hand side of (\ref{limit_spectral}) is non-negative, see Fig.\,\ref{fig:tangens}.

\section*{Appendix A: The reduction of the general case to the one treated in Section \ref{sect:asymp_diag}}\label{App_Kis}

We proceed as follows. First, we decompose the graph $\widehat{\mathbb G}$ into the union of its stiff and soft components, $\widehat{\mathbb G}=\mathbb G^{\text{soft}}\cup \mathbb G^{\text{stiff}}$, each of these being a graph on its own. The common boundary of them is $\partial \mathbb G:=\mathbb G^{\text{soft}}\cap \mathbb G^{\text{stiff}},$ and it is treated as a set of vertices. Second, we consider two maximal operators $\breve A_{\max}^{\text{soft}}$ and $\breve A_{\max}^{\text{stiff}},$ which are densely defined in $L_2(\mathbb G^{\text{soft}})$ and $L_2(\mathbb G^{\text{stiff}})$, respectively, by \eqref{diff_expr}, \eqref{domAmax} applied to $\mathbb G^{\text{soft}}$ and $\mathbb G^{\text{stiff}}$. Furthermore, we introduce the orthogonal projections $P^{\text{soft}}, P^{\text{stiff}}$ in the boundary space $\mathcal H$ onto the subspaces pertaining to vertices of $\mathbb G^{\text{soft}}$ and $\mathbb G^{\text{stiff}}$, respectively. Finally, we construct boundary triples for $\breve A_{\max}^{\text{soft (stiff)}}$  with boundary spaces $P^{\text{soft (stiff)}}\mathcal H$ and boundary operators $\breve\Gamma_j^{\text{soft (stiff)}}$, $j=0,1$ ({\it cf.} \eqref{boundary_operators}), respectively.

Now consider the restrictions
\begin{equation}
\begin{aligned}
&A_{\max}^{\text{soft (stiff)}}=\breve A_{\max}^{\text{soft (stiff)}}\big|_{\dom(A_{\max}^{\text{soft (stiff)}})},\\[0.4em]
&\dom\bigl(A_{\max}^{\text{soft (stiff)}}\bigr):=\Bigl\{u\in \dom\bigl(\breve A_{\max}^{\text{soft (stiff)}}\bigr)\Big| (1-P_{\partial \mathbb G})\breve\Gamma_1^{\text{soft (stiff)}}u=0\Bigr\},
\end{aligned}
\label{maternoe_slovo}
\end{equation}
where $P_{\partial \mathbb G}$ is defined as an orthogonal projection in $\mathcal H$ onto the subspace pertaining to the vertices belonging to $\partial \mathbb G$. For these two maximal operators, one has the common boundary space $P_{\partial\mathbb G}\mathcal H$ and boundary operators defined by
$$
\Gamma_j^{\text{soft (stiff)}}:=P_{\partial\mathbb G} \breve\Gamma_j^{\text{soft (stiff)}},\quad j=0,1.
$$
The corresponding $M$-matrices $M^{\text{soft (stiff)}}$ are computed as inverses of the matrices $$P_{\partial \mathbb G}\bigl(\breve M^{\text{soft (stiff)}}\bigr)^{-1}P_{\partial \mathbb G},$$ where the latter are considered in the reduced space $P_{\partial \mathbb G} \mathcal H$ and $\breve M^{\text{soft (stiff)}}$ are $M$-matrices of $\breve A_{\max}^{\text{soft (stiff)}}$ relative to the boundary triples
$\bigl(P^{\text{soft (stiff)}}\mathcal H, \breve\Gamma_0^{\text{soft (stiff)}}, \breve\Gamma_1^{\text{soft (stiff)}}\bigr)$.

It is easily shown that the operator $A^\e_t$ is expressed as an almost solvable extension parameterised by the matrix $B=0$ relative to a triple which has the $M$-matrix $M=M^{\text{soft}}+M^{\stiff}$. It follows that all the prerequisites of the analysis carried out in Section \ref{sect:asymp_diag} are met.

\section*{Appendix B: Proof of Lemma \ref{mu_lemma}}


The proof could be carried out on the basis of \cite{Yorzh3}, \cite{Yorzh4} and is rather elementary. Nevertheless, in the present paper we have elected to follow an alternative approach to this proof, which has an advantage of carrying over to the PDE case with minor modifications.

For simplicity we set $w_V(e)=1$ for all $e, V$ in (\ref{Atau1}), as the argument below is unaffected by the concrete choice of the list $\{w_V(e)\}_{e\ni V},$ $V\in\widehat{\mathbb G},$ in the construction of Section \ref{Gelfand_section}. For convenience, we also imply that the unitary rescaling to a graph of length one has been applied to the  operator family $A_t^\varepsilon$. For brevity, we keep the same notation for the unitary images of graphs $\widehat{\mathbb{G}}$, $\mathbb{G}^{\rm stiff}$ and $\partial \mathbb G$ under this transform.

For each $\tau\in[-\pi, \pi),$ the eigenvalues of $B_0(\tau)$ are those $\mu\in{\mathbb C}$ for which there exists $u\neq0$ satisfying
\begin{equation}
\left\{\begin{array}{ll}\biggl(\dfrac{d}{dx}+{\rm i}\tau\biggr)^2u=0\quad{\rm in}\ {\mathbb G}^{\rm stiff}, \\[1.2em]
-\sum_{e\ni V}\sigma_e\bigl(u'_{e}(V)+{\rm i}\tau u(V)\bigr)=\mu u(V),\quad V\in\partial{\mathbb G},\\[1em]
u\ {\rm continuous\ on\ }{\mathbb G}^{\rm stiff},
\end{array}\right.
\label{problem}
\end{equation}
where $u'_{e}(V)$ is the derivative of $u$ along the edge $e$ of ${\mathbb G}^{\rm stiff}$ evaluated at $V\in\partial{\mathbb G},$ and, as before,
$\sigma_{e}=-1$ or $\sigma_{e}=1,$ depending on whether $e$ is  incoming or outgoing for $V,$ respectively.
It is known that the spectrum of (\ref{problem}) is discrete and the least eigenvalue, which clearly coincides with $\mu^{(\tau)},$ is simple.

{\it Formal series.}  In order to show (\ref{mu_asymptotics}), we first consider  series in powers of ${\rm i}\tau:$
\begin{equation}
\mu=\sum_{k=1}^\infty\alpha_j({\rm i}\tau)^{2k},\qquad
  u=\sum_{j=0}^\infty u_j({\rm i}\tau)^j,
  \label{expansion}
\end{equation}
where $u_j,$ $j=1,2,\dots$ are continuous on ${\mathbb G}^{\rm stiff}.$

Note that the expansion for $\mu$ contains only even powers of the parameter $\tau,$ as it is an even function of $\tau.$ Indeed, the function obtained from the eigenfunction $u$ in (\ref{problem}) by changing the directions of all edges of the graph is clearly an eigenfunction for (\ref{problem}) with $\tau$  replaced by $-\tau.$ (On such a change of edge direction, the weights $w_e(V),$ ${e\ni V},$ $V\in\widehat{\mathbb G},$ are replaced by their complex conjugates.) In view of the fact that for all $\tau\in(-\pi,\pi]$ the eigenvalue $\mu^{(\tau)}$ is simple, we obtain $\mu^{(-\tau)}=\mu^{(\tau)}.$

Substituting the expansion (\ref{expansion}) into (\ref{problem}) and equating the coefficients on different powers of
$\tau,$ we obtain a sequence of recurrence relations for $u_j,$ $j=0,1,\dots$  In particular, the problem for $u_0$ is obtained by comparing the coefficients on $\tau^0:$
\[
\left\{\begin{array}{lll}u_0''=0\ \ \ {\rm on}\ \ {\mathbb G}^{\rm stiff}, \\[0.5em]
\sum_{e\ni V}\sigma_e(u_0)_e'(V)=0,\quad V\in\partial{\mathbb G},\\[0.6em]
u_0\ {\rm continuous\ on\ }{\mathbb G}^{\rm stiff}.

\end{array}\right.
\]
Assuming that ${\mathbb G}^{\rm stiff}$ contains a loop, it follows that
$u_0$ is a constant, which we set to be unity. In the case opposite, i.e., when  ${\mathbb G}^{\rm stiff}$ is a tree, $\mu^{(\tau)}\equiv 0$  for all $\tau$,  and the claim of Lemma follows trivially.

We impose the condition of vanishing mean of $u_j,$ $j=1,2,\dots$ over ${\mathbb G}^\stiff.$ This is justified by the convergence estimates below as well as the fact that the eigenvalue $\mu$ is simple. The choice $u_0=1$ thus corresponds to the ``normalisation" condition that the mean over ${\mathbb G}^\stiff$ of the eigenfunction $u$ for (\ref{problem}) is close to unity\footnote{The eigenfunction $u$ clearly does not vanish identically, at least for small values of $\tau.$} for small values of $\tau.$


Proceeding with the asymptotic procedure, the problem for $u_1$ is obtained by comparing the coefficients on $\tau^1:$
\[
\left\{\begin{array}{ll}u''_1=0\ \ {\rm on}\ \ {\mathbb G}^{\rm stiff}, \\[0.7em]
\sum_{e\ni V}\sigma_e\bigl((u_1)_e'(V)+1\bigr)=0,\quad V\in\partial{\mathbb G},\\[0.8em]
u_1\ {\rm continuous\ on\ }{\mathbb G}^{\rm stiff},\\[0.8em]
\int_{{\mathbb G}^{\rm stiff}}u_1=0.
\end{array}\right.
\]
Further, the equation for $u_2$ is obtained by comparing the coefficients on  $\tau^2:$
\begin{equation}
\left\{\begin{array}{ll}u''_2=-2u'_1-1 \ \ {\rm on}\ \ {\mathbb G}^{\rm stiff}, \\[1.1em]
-\sum_{e\ni V}\sigma_e\bigl((u_2)_e'(V)+u_1(V)\bigr)=\alpha_2,\quad V\in\partial{\mathbb G},\\[1.2em]
u_2\ {\rm continuous\ on\ }{\mathbb G}^{\rm stiff},\\[0.8em]
\int_{{\mathbb G}^{\rm stiff}}u_2=0.
\end{array}\right.
\label{u_2}
\end{equation}
 The condition for solvability of the problem (\ref{u_2}) yields the expression for $\alpha_2,$ as follows:
\[
\int_{{\mathbb G}^{\rm stiff}}(-2u'_1-1)=\int_{{\mathbb G}^{\rm stiff}}u''_2=-\sum_{V\in\partial{\mathbb G}}\ \sum_{e\ni V}\sigma_e(u_2)_e'(V)
=\sum_{V\in\partial{\mathbb G}}\Bigl(\sum_{e\ni V}\sigma_e u_1(V)+\alpha_2\Bigr).
\]
Re-arranging the terms in the last equation, we obtain
\[
\alpha_2=-\bigl\vert\partial{\mathbb G}\bigr\vert^{-1}\int_{{\mathbb G}^{\rm stiff}}(u'_1+1).
\]
The above asymptotic procedure is continued, to obtain the terms of all orders in (\ref{expansion}). In particular, for the term $u_3$ in the expansion for $u$ we obtain
\begin{equation*}
\left\{\begin{array}{ll}u''_3=-2u'_2-u_1 \ \ {\rm on}\ \ {\mathbb G}^{\rm stiff}, \\[1.1em]
-\sum_{e\ni V}\sigma_e\bigl((u_3)_e'(V)+u_2(V)\bigr)=\alpha_2u_1,\quad V\in\partial{\mathbb G},\\[1.2em]
u_3\ {\rm continuous\ on\ }{\mathbb G}^{\rm stiff},\\[0.8em]
\int_{{\mathbb G}^{\rm stiff}}u_3=0.
\end{array}\right.
\label{u_3}
\end{equation*}

{\it Error estimates.}
We write
\[
u=1+{\rm i}\tau u_1+({\rm i}\tau)^2u_2+({\rm i}\tau)^3u_3+R,\qquad \mu^{(\tau)}=\alpha_2({\rm i}\tau)^2+r,
\]
so that $R,$ $r$ satisfy
\begin{empheq}[right=\empheqrbrace]{align}
&\biggl(\dfrac{d}{dx}+{\rm i}\tau\biggr)^2R=-({\rm i}\tau)^4(2u_3'+u_2)-({\rm i}\tau)^5u_3\quad \text{ on } \mathbb{G}^{\rm stiff},\label{R_equation}
\\[0.4em]
&-\sum_{e\ni V}\sigma_e (R'_{e}(V)+{\rm i}\tau R(V))=\label{bc}\\
&=\bigl(r+\alpha_2({\rm i}\tau)^2\bigr)
\bigl(1+{\rm i}\tau u_1+({\rm i}\tau)^2u_2+({\rm i}\tau)^3u_3+R\bigr)\nonumber\\[0.5em]
&-\alpha_2({\rm i}\tau)^2(1+{\rm i}\tau u_1),\quad V\in\partial \mathbb{G}\nonumber\\[0.4em]
&R\ {\rm continuous\ on\ }{\mathbb G}^{\rm stiff},\nonumber\\[0.4em]
&\int_{{\mathbb G}^{\rm stiff}}R=0.\ \ \ \ \ \ &\nonumber
\end{empheq}

Notice first that
\begin{multline}
r+\alpha_2({\rm i}\tau)^2=\mu^{(\tau)}=\min_{u\in W^{2,2}({\mathbb G}^{\rm stiff})}\biggl(\sum_{\partial{\mathbb G}}\vert u\vert^2\biggr)^{-1}\int_{{\mathbb G}^{\rm stiff}}\Biggl\vert\biggl(\dfrac{d}{dx}+{\rm i}\tau\biggr)u\Biggr\vert^2 \le\bigl\vert\partial{\mathbb G}\bigr\vert^{-1}\bigl\vert {\mathbb G}^{\rm stiff}\bigr\vert\tau^2.
\label{mu_est}
\end{multline}
Multiplying (\ref{R_equation}) by $R$, integrating by parts, and using (\ref{bc}), we obtain the estimate
\begin{equation}
\Vert R\Vert_{L^2({\mathbb G}^{\rm stiff})}^2\le C\bigl(\vert\tau\vert\vert r\vert\Vert R\Vert_{L^2({\mathbb G}^{\rm stiff})}+\vert\tau\vert^4\Vert R\Vert_{L^2({\mathbb G}^{\rm stiff})}+\vert r\vert^2\bigr),\qquad C>0,
\label{R_est}
\end{equation}
and hence, by virtue of (\ref{mu_est}), we obtain
\begin{equation}
\Vert R\Vert_{L^2({\mathbb G}^{\rm stiff})}\le C\tau^2.
\label{first_R_estimate}
\end{equation}

Next, we re-arrange the right-hand side of (\ref{bc}):
\begin{multline*}
\bigl(r+\alpha_2({\rm i}\tau)^2\bigr)
\bigl(1+{\rm i}\tau u_1+({\rm i}\tau)^2u_2+({\rm i}\tau)^3u_3+R\bigr)-\alpha_2({\rm i}\tau)^2(1+{\rm i}\tau u_1)\\[0.5em]=r\bigl(1+{\rm i}\tau u_1+({\rm i}\tau)^2u_2+({\rm i}\tau)^3u_3+R\bigr)+\alpha_2({\rm i}\tau)^2\bigl(({\rm i}\tau)^2u_2+({\rm i}\tau)^3u_3+R\bigr).
\end{multline*}
Multiplying  (\ref{R_equation}) by $1$, integrating by parts, and using (\ref{bc}) once again yields the existence of  $C>0$ such that
\begin{equation}
\vert r\vert\le C\bigl(\vert\tau\vert\Vert R\Vert_{L^2({\mathbb G}^{\rm stiff})}+\vert\tau\vert^4\bigr).
\label{r_estimate}
\end{equation}
Combining this with (\ref{first_R_estimate}) yields $\vert r\vert\le C\tau^3,$ which, by virtue of (\ref{R_est}) again, implies
\begin{equation}
\Vert R\Vert_{L^2({\mathbb G}^{\rm stiff})}\le C\vert\tau\vert^3.
\label{second_R_estimate}
\end{equation}
Finally, the inequalities (\ref{r_estimate}) and (\ref{second_R_estimate}) together yield
\begin{equation}
\vert r\vert\le C|\tau|^4,
\label{r_final}
\end{equation}
as claimed.\footnote{Combining (\ref{r_final}) with (\ref{mu_est}), we also obtain the estimate $\Vert R\Vert_{L^2({\mathbb G}^{\rm stiff})}\le C\tau^4.$}



\section*{Appendix C: Proof of Lemma \ref{bound_below_lemma}}

For all $\tau\in[-\pi,\pi),$ using the formula for the second eigenvalue $\mu_2^{(\tau)}$ of the problem (\ref{problem}) via the Rayleigh quotient, we obtain
\begin{align*}
\mu_2^{(\tau)}&=\min\Biggl\{\biggl(\sum_{\partial{\mathbb G}}\vert u\vert^2\biggr)^{-1}\int_{{\mathbb G}^{\rm stiff}}\Biggl\vert\biggl(\dfrac{d}{dx}+{\rm i}\tau\biggr)u\Biggr\vert^2: u\in W^{2,2}({\mathbb G}^{\rm stiff}), \int_{{\mathbb G}^{\rm stiff}}u=0\Biggr\}\\[0.9em]
&\ge \min\Biggl\{\biggl(\sum_{\partial{\mathbb G}}\vert u\vert^2\biggr)^{-1}\int_{{\mathbb G}^{\rm stiff}}\vert u'\vert^2: u\in W^{2,2}({\mathbb G}^{\rm stiff}), \int_{{\mathbb G}^{\rm stiff}}u=0\Biggr\}=\mu_2^{(0)}>0,
\end{align*}
from which the claim follows by setting $C_\perp=\mu_2^{(0)}.$

\section*{Acknowledgements}

We are grateful to Professor S. Naboko for suggesting a calculation in Section 8.

\bibliographystyle{siamplain}

\end{document}